\documentclass[journal]{IEEEtran} %
\usepackage{amsmath, amsfonts, amsthm} %
\usepackage{algorithmic}
\usepackage{algorithm}
\usepackage[caption=false, font=normalsize, labelfont=sf, textfont=sf]{subfig}
\usepackage{url}
\usepackage{graphicx}

\usepackage{bm}
\usepackage{tikz}
\usepackage{pgfplots}
\pgfplotsset{compat=newest}
\usetikzlibrary{calc, shapes.geometric, shapes.misc, arrows, matrix, fit, decorations.markings}
\usepackage[absolute,overlay]{textpos}

\newcommand{\aprob}{\ensuremath{q}} 				%
\newcommand{\param}{\ensuremath{\theta}} 			%
\newcommand{\params}{\ensuremath{\bog{\param}}} 	%
\newcommand{\Nsamp}{\ensuremath{N}} 				%
\newcommand{\idxi}{\ensuremath i} %

\newcommand{\bo}[1]{\ensuremath{\mathbf{#1}}}
\newcommand{\bog}[1]{\ensuremath{\bm{#1}}}
\newcommand{\txt}[1]{\ensuremath{\textrm{#1}}}

\newcommand{\st}{\ensuremath{\quad \txt{s.t.} \quad}} %
\newcommand{\eqpoint}{\ensuremath{\,.}} %

\DeclareMathOperator*{\armax}{arg\,max}

\newcommand{\argmax}[1][\cdot]{\ensuremath{\underset{#1}{\armax}\ }}

\newcommand{\lpa}{\ensuremath\left(}
\newcommand{\rpa}{\ensuremath\right)}
\newcommand{\lbb}{\ensuremath\left[}
\newcommand{\rbb}{\ensuremath\right]}
\newcommand{\lcb}{\ensuremath\left\{}
\newcommand{\rcb}{\ensuremath\right\}}

\newcommand{\summ}[2][]{\ensuremath{\sum \limits_{#2}^{#1}}} %

\DeclareMathOperator{\E}{E} %
\newcommand{\eval}[2][\cdot]{\ensuremath \E_{#1}\negthinspace\lbb#2\rbb} 	%
\newcommand{\evalnb}[2][\cdot]{\ensuremath \E_{#1}\negthinspace[#2]} 		%
\newcommand{\dkl}[2]{\ensuremath D_{\text{KL}}\lpa#1 \parallel #2\rpa} 		%
\newcommand{\he}{\mathcal{H}}												%
\newcommand{\HE}[1][\cdot]{\ensuremath \he\lpa#1\rpa} 						%

\newcommand{\rv}{RV}
\newcommand{\Nb}{\ensuremath{N_{\txt{b}}}}                  %
\newcommand{\Ne}{\ensuremath{N_{\txt{e}}}}                  %
\newcommand{\stddev}{\ensuremath{\sigma}}                   %
\newcommand{\noisestd}{\ensuremath{\stddev_{\txt{n}}}}      %
\newcommand{\noisevar}{\ensuremath{n}}                      %
\newcommand{\noisevec}{\ensuremath{\bo{\noisevar}}}         %
\newcommand{\Nclass}{\ensuremath{M}}                        %
\newcommand{\xvar}{\ensuremath{x}}                          %
\newcommand{\xvec}{\ensuremath{\bo{x}}}                     %
\newcommand{\yvar}{\ensuremath{y}}                          %
\newcommand{\yvec}{\ensuremath{\bo{\yvar}}}                 %
\newcommand{\prob}{\ensuremath{p}}                          %
\newcommand{\indx}{\ensuremath{n}}                          %
\newcommand{\indi}{\ensuremath{i}}                          %
\newcommand{\relu}[1][\cdot]{\ensuremath{\rho_{1} \lpa #1 \rpa}}%

\newcommand{\semvar}{\ensuremath{s}} 							%
\newcommand{\semvarreal}{\ensuremath{\mathsf{\semvar}}} 		%
\newcommand{\semvec}{\ensuremath{\bo{\semvar}}} 				%
\newcommand{\semvecreal}{\ensuremath{\bog{\semvarreal}}} 		%
\newcommand{\semhvec}{\ensuremath{\hat{\semvec}}} 				%
\newcommand{\rate}{\ensuremath R} 								%
\newcommand{\rc}{\ensuremath \rate_{\text{C}}} 					%
\newcommand{\rcind}[1][]{\ensuremath \ifthenelse{\equal{#1}{}}{R_{\text{C}}}{R_{\text{C},#1}}} %
\newcommand{\capa}{\ensuremath C} 								%
\newcommand{\MI}{\ensuremath I} 								%
\newcommand{\mi}[2][]{\ensuremath \MI_{#1}\lpa#2\rpa} 			%
\newcommand{\Lossf}{\ensuremath \mathcal{L}} 					%
\newcommand{\mic}{\ensuremath \MI_{\txt{C}}} 					%
\newcommand{\relvar}{\ensuremath{z}} 							%
\newcommand{\relvarreal}{\ensuremath{\mathsf{z}}} 				%
\newcommand{\relvarest}{\ensuremath{\hat{\relvar}}} 			%
\newcommand{\relvecest}{\ensuremath{\hat{\bo{\relvar}}}} 		%
\newcommand{\relvec}{\ensuremath{\bo{\relvar}}} 				%
\newcommand{\relvecreal}{\ensuremath{\bog{\relvarreal}}} 		%
\newcommand{\featvar}{\ensuremath{r}} 							%
\newcommand{\featvec}{\ensuremath{\bo{\featvar}}} 				%
\newcommand{\formlang}{\ensuremath{L}}							%
\newcommand{\lagrange}{\ensuremath{\beta}} 						%
\newcommand{\ntx}{\ensuremath{N_{\txt{Tx}}}} 					%
\newcommand{\nrx}{\ensuremath{N_{\txt{w}}}} 					%
\newcommand{\Nrx}{\ensuremath{N_{\txt{Rx}}}} 					%
\newcommand{\Nfeat}{\ensuremath{N_{\txt{Feat}}}} 				%
\newcommand{\speceff}{\ensuremath{\eta}} 						%
\newcommand{\centralclass}{Central} 							%
\newcommand{\sinfoni}{SINFONY} 									%
\newcommand{\txrx}{Tx/Rx} 					                    %
\newcommand{\sinfonitxrx}{\sinfoni\ - \txrx} 					%
\newcommand{\perfcomm}{Perfect com.} 		                    %
\newcommand{\sinfoniperfcomm}{\sinfoni\ - \perfcomm} 		    %
\newcommand{\sinfonichan}{\sinfoni\ - AWGN} 					%
\newcommand{\sinfonichantrain}{\sinfoni\ - AWGN + training} 	%
\newcommand{\sinfoniclassdig}{\sinfoni\ - Classic digital com.} 	%
\newcommand{\sinfoniclassanae}{\sinfoni\ - Analog semantic AE} 	%
\newcommand{\centralimagecomm}{\centralclass\ image com.} 	    %
\newcommand{\Nsemvec}{\ensuremath{N_{\semvar}}} 				%
\newcommand{\Nrelvec}{\ensuremath{N_{\relvar}}} 				%
\newcommand{\Nimx}{\ensuremath{N_{\txt{x}}}} 					%
\newcommand{\Nimy}{\ensuremath{N_{\txt{y}}}} 					%
\newcommand{\Nimc}{\ensuremath{N_{\txt{c}}}} 					%
\newcommand{\setsym}{\ensuremath{M}} 							%
\newcommand{\dom}{\ensuremath{\mathcal{\setsym}}} 				%
\newcommand{\semset}{\ensuremath{\dom_{\semvar}}} 				%
\newcommand{\relset}{\ensuremath{\dom_{\relvar}}} 				%
\newcommand{\xset}{\ensuremath{\dom_{\xvar}}} 					%
\newcommand{\yset}{\ensuremath{\dom_{\yvar}}} 					%

\newcommand{\txpars}{\ensuremath{\params}} 						%
\newcommand{\Ntxpars}{\ensuremath{N_{\txpars}}} 				%
\newcommand{\Nrxpars}{\ensuremath{N_{\rxpars}}} 				%
\newcommand{\rxpar}{\ensuremath{\varphi}} 						%
\newcommand{\rxpars}{\ensuremath{\bog{\rxpar}}} 				%
\newcommand{\varpar}{\ensuremath{\vartheta}} 					%
\newcommand{\varpars}{\ensuremath{\bog{\varpar}}} 				%
\newcommand{\covariance}{\ensuremath{\bog{\Sigma}}} 			%
\newcommand{\mean}{\ensuremath{\mu}} 							%
\newcommand{\funcrtrick}{\ensuremath{f}} 						%
\newcommand{\meanfunc}{\ensuremath{\mean}} 						%
\newcommand{\dirac}{\ensuremath{\delta}} 						%
\newcommand{\ceparams}{\ensuremath{\Lossf_{\txpars, \rxpars}^{\txt{CE}}}}	%
\newcommand{\encdet}{\ensuremath{\meanfunc_{\txpars}(\semvec)}} 		%
\newcommand{\lr}{\ensuremath{\epsilon}}                                 %
\newcommand{\xvarreal}{\ensuremath{\mathsf{\xvar}}} 		            %
\newcommand{\xvecreal}{\ensuremath{\bog{\xvarreal}}} 		            %
\newcommand{\yvarreal}{\ensuremath{\mathsf{\yvar}}} 		            %
\newcommand{\yvecreal}{\ensuremath{\bog{\yvarreal}}} 		            %
\newcommand{\noisevarreal}{\ensuremath{\mathsf{\noisevar}}}             %
\newcommand{\noisevecreal}{\ensuremath{\bog{\noisevarreal}}}            %
\newcommand{\snr}{\ensuremath{\txt{SNR}}}                               %
\newcommand{\reinforce}{REINFORCE} 		                                %
\newcommand{\xvecwnorm}{\ensuremath{\tilde{\xvec}}} 		            %

\newcommand{\abs}[1][\cdot]{\ensuremath{\left|#1\right|}}

\newcommand{\norm}[1]{\ensuremath{\left\|#1\right\|}}
\newcommand{\lnorm}[1][\cdot]{\ensuremath{l_{#1}}}

\newcommand{\trapo}{\ensuremath{T}} %
\newcommand{\realnum}{\ensuremath{\mathbb{R}}} %

\newcommand{\idxC}{\ensuremath k} %

\newcommand{\mF}[1][\nC]{\ensuremath \mathbf{F}_{\nC}} %

\newcommand{\eA}{\ensuremath a} %
\newcommand{\setA}[1][]{\ensuremath\mathcal{\MakeUppercase{\eA}}_{#1}} %
\newcommand{\defD}[1][0]{\ensuremath\lcb\vdwir[\idxC]\,\vert\,\sd_{\idxC,\idxA}\in\ifthenelse{#1>0}{\setA[\idxC,\idxA]}{\setA}\quad\forall\,\idxA=1,\dotsc,2\nA\rcb} %

\newcommand{\nA}{\ensuremath N_{\text{L}}} %
\newcommand{\nC}{\ensuremath N_{\text{C}}} %

\newcommand{\vrc}[1][]{\ensuremath \ifthenelse{\equal{#1}{}}{\mathbf{r}_{\text{C}}}{\mathbf{r}_{\text{C},#1}}} %
\newcommand{\rce}[1][]{\ensuremath \ifthenelse{\equal{#1}{}}{\hat{R}_{\text{C}}}{\hat{R}_{\text{C},#1}}} %

\newcommand{\drc}[1][]{\ensuremath \ifthenelse{\equal{#1}{}}{R'_{\text{C}}}{R'_{\text{C},#1}}} %

\newcommand{\sd}{\ensuremath d} %

\usepackage{pgfplots}
\pgfplotsset{compat=newest}
\usetikzlibrary{pgfplots.groupplots}
\usetikzlibrary{calc,shapes.geometric,shapes.misc,arrows,matrix}
\usepackage[absolute,overlay]{textpos} %

\definecolor{darkgreen}{rgb}{0,0.498039,0}
\definecolor{darkyellow}{rgb}{0.75,0.75,0}
\definecolor{magenta}{rgb}{0.75,0,0.75}
\newcommand{\clsd}{darkgray} %
\newcommand{\clmmse}{darkgray} %
\newcommand{\clmosic}{purple} %
\newcommand{\clamp}{darkgreen} %
\newcommand{\clsdr}{orange} %
\newcommand{\clmmnet}{darkyellow} %
\newcommand{\cldetnet}{magenta} %
\newcommand{\cloamp}{blue} %
\newcommand{\clcmd}{red} %
\newcommand{\clcmdshal}{orange} %
\newcommand{\clhcmd}{darkgreen} %
\newcommand{\clawgn}{green!50!black} %

\pgfplotsset{
	ptsd/.style={color=\clsd, mark=*, mark size=1.4
	}, %
	ptmmse/.style={color=\clmmse, mark=triangle*, dashed, mark options={solid}, mark size=1.4
	}, %
	ptmosic/.style={color=\clmosic, mark=triangle*, mark size=1.4
	}, %
	ptamp/.style={color=\clamp, mark=diamond*, mark size=1.4
	}, %
	ptsdr/.style={color=\clsdr, mark=*, dashed, mark options={solid}, mark size=1.4
	}, %
	ptdetnet/.style={color=\cldetnet, mark=square*, dashed, mark options={solid}, mark size=1.4
	}, %
	ptmmnet/.style={color=\clmmnet, mark=diamond*, dashed, mark options={solid}, mark size=1.4
	}, %
	ptoamp/.style={color = \cloamp, mark=x, mark size=4*0.7, mark options={solid}
	}, %
	ptcmd/.style={color=\clcmd, mark=square*, mark size=1.4
	}, %
	ptcmdshal/.style={\clcmdshal, dashed, mark=square*, mark options={solid}, mark size=2*0.7
	}, %
	pthcmd/.style={color=\clhcmd, mark size=1.4, , mark options={solid}, %
	}, %
	ptawgn/.style={color=\clawgn, semithick
	}, %
}

\pgfplotsset{
	every axis plot/.style={line width=1.4pt},
	tick pos=left,
	yminorgrids,
	every outer x axis line/.style={line width=1.4pt},
	every outer y axis line/.style={line width=1.4pt},
	every tick/.style={
		black,
		line width=0.7pt,
	},
	label style={font=\fontsize{8}{9}\selectfont},
	title style={font=\fontsize{8}{9}\selectfont},
	every y tick label/.style={font=\fontsize{8}{9}\color{black}},
	every x tick label/.style={font=\fontsize{8}{9}\color{black}},
	every extra x tick/.style={grid style={solid, violet, line width=2.8pt},
		x tick label style={/pgf/number format/.cd,precision=10}
	},
	legend style= {
		line width=0.7pt,
		font=\fontsize{11.4}{12}\selectfont\color{black},
		legend cell align=left,
		align=left,
		fill=white,
		fill opacity=0.8, draw opacity=1, text opacity=1,
		nodes={scale=0.7, transform shape},
	},
	height = 8cm, %
}

\usepackage{scalerel}
\usetikzlibrary{svg.path}
\definecolor{orcidlogocol}{HTML}{A6CE39}
\tikzset{
	orcidlogo/.pic={
		\fill[orcidlogocol] svg{M256,128c0,70.7-57.3,128-128,128C57.3,256,0,198.7,0,128C0,57.3,57.3,0,128,0C198.7,0,256,57.3,256,128z};
		\fill[white] svg{M86.3,186.2H70.9V79.1h15.4v48.4V186.2z}
		svg{M108.9,79.1h41.6c39.6,0,57,28.3,57,53.6c0,27.5-21.5,53.6-56.8,53.6h-41.8V79.1z M124.3,172.4h24.5c34.9,0,42.9-26.5,42.9-39.7c0-21.5-13.7-39.7-43.7-39.7h-23.7V172.4z}
		svg{M88.7,56.8c0,5.5-4.5,10.1-10.1,10.1c-5.6,0-10.1-4.6-10.1-10.1c0-5.6,4.5-10.1,10.1-10.1C84.2,46.7,88.7,51.3,88.7,56.8z};
	}
}

\newcommand\orcidicon[1]{\href{https://orcid.org/#1}{\mbox{\scalerel*{
				\begin{tikzpicture}[yscale=-1,transform shape]
					\pic{orcidlogo};
				\end{tikzpicture}
			}{U}}}} %
\usepackage{hyperref}

\begin{document}

\title{Semantic Information Recovery in Wireless Networks} %

\author{Edgar~Beck$^{\orcidicon{0000-0003-2213-9727}}$,~\IEEEmembership{Graduate~Student~Member,~IEEE,}
	Carsten~Bockelmann$^{\orcidicon{0000-0002-8501-7324}}$,~\IEEEmembership{Member,~IEEE,}
	and~Armin~Dekorsy$^{\orcidicon{0000-0002-5790-1470}}$,~\IEEEmembership{Senior~Member,~IEEE}%
\thanks{This work was partly funded by the Federal State of Bremen and the University of Bremen as part of the Human on Mars Initiative, and by the German Ministry of Education and Research (BMBF) under grant 16KISK016 (Open6GHub).}%
\thanks{The authors are with the Gauss-Olbers Space Technology Transfer Center c/o Department of Communications Engineering, University of Bremen, 28359 Bremen, Germany (e-mail: \{beck, bockelmann, dekorsy\}@ant.uni-bremen.de).}%
}

\maketitle

\begin{abstract}
	Motivated by the recent success of Machine Learning (ML) tools in wireless communications, the idea of semantic communication by Weaver from 1949 has gained attention. It breaks with Shannon's classic design paradigm by aiming to transmit the meaning of a message, i.e., semantics, rather than its exact version and thus allows for savings in information rate.
	In this work, we extend the fundamental approach from Basu et al. for modeling semantics to the complete communications Markov chain. Thus, we model semantics by means of hidden random variables and define the semantic communication task as the data-reduced and reliable transmission of messages over a communication channel such that semantics is best preserved. We cast this task as an end-to-end Information Bottleneck problem, allowing for compression while preserving relevant information most. As a solution approach, we propose the ML-based semantic communication system \sinfoni\ and use it for a distributed multipoint scenario: \sinfoni\ communicates the meaning behind multiple messages that are observed at different senders to a single receiver for semantic recovery. We analyze \sinfoni\ by processing images as message examples. Numerical results reveal a tremendous rate-normalized SNR shift up to 20 dB compared to classically designed communication systems.
\end{abstract}


\section{Introduction}

\IEEEPARstart{W}{hen} Shannon laid the theoretical foundation of the research area of communications engineering back in 1948, he deliberately excluded semantic aspects from the system design \cite{shannon_mathematical_1948, weaver_recent_1949}. In fact, the idea of addressing semantics in communications arose shortly after Shannon's work in \cite{weaver_recent_1949}, but it remained largely unexplored. Since then, the design focus of communication systems has been on digital error-free symbol transmission. In recent years, it has become clear that semantics-agnostic communication limits the achievable efficiency in terms of bandwidth, power, and complexity trade-offs. Notable examples include wireless sensor networks and broadcast scenarios \cite{gastpar_code_2003}.

Owing to the great success of Artificial Intelligence (AI) and, in particular, its subdomain Machine Learning (ML), %
ML tools have been recently investigated for wireless communications \cite{oshea_introduction_2017, simeone2018very, beck_cmdnet_2021}. Now, ML with its ability to extract features appears to be a proper means to realize a semantic design. Further, we note that the latter design is supported and possibly enabled by the 6G vision of integrating AI and ML on all layers of the communications system design, i.e., by an ML-native air interface.

Motivated by these new ML tools and driven by unprecedented needs of the next wireless communication standard 6G in terms of data rate, latency, and power, the idea of semantic communication has received considerable attention \cite{weaver_recent_1949, popovski_semantic-effectiveness_2020, calvanese_strinati_6g_2021, lan_what_2021, uysal_semantic_2022, gunduz_beyond_2023}. It breaks with the existing classic design paradigms by including semantics in the design of the wireless transmission. The goal of such a transmission is, therefore, to deliver the required data from which the highest levels of quality of information may be derived, as perceived by the application and/or the user. More precisely, semantic communication aims to transmit the meaning of a message rather than its exact version and hence allows for compression and coding to the actual semantic content. Thus, savings in bandwidth, power, and complexity are expected.

\subsection{Related Work}
\label{sec:1_1}

The notion of semantic communication traces back to Weaver \cite{weaver_recent_1949} who reviewed Shannon's information theory \cite{shannon_mathematical_1948} in 1949 and amended considerations w.r.t. semantic content of messages. Oftentimes quoted is his statement that \textit{``there \textbf{seem} to be [communication] problems at three levels"} \cite{weaver_recent_1949}:
\begin{itemize}
	\item[A.] How accurately can the symbols of communication be transmitted? (The technical problem.)
	\item[B.] How precisely do the transmitted symbols convey the desired meaning? (The semantic problem.)
	\item[C.] How effectively does the received meaning affect conduct in the desired way? (The effectiveness problem.)
\end{itemize}
Since then semantic communication was mainly investigated from a philosophical point of view, see, e.g., \cite{floridi_philosophical_2009, hofkirchner_emergent_2013}.

The generic model of Weaver was revisited by Bao, Basu et al. in \cite{bao_towards_2011, basu_preserving_2014} where the authors define semantic information source and semantic channel. In particular, the authors consider a semantic source that \textit{``observes the world and generates meaningful messages characterizing these observations"} \cite{basu_preserving_2014}. The source is equivalent to conclusions, i.e., ``models" of the world, that are unequivocally drawn following a set of known inference rules based on observation of messages. In \cite{bao_towards_2011}, the authors consider joint semantic compression and channel coding at Level B with the classic transmission system, i.e., Level A, as the (semantic) channel. In contrast, \cite{basu_preserving_2014} only deals with semantic compression and uses a different definition of the semantic channel (which we will make use of in this article): It is equal to the entailment relations between ``models" and ``messages". By this means, the authors are able to derive semantic counterparts of the source and channel coding theorems. However, as the authors admit, these theorems do not tell how to develop optimal coding algorithms and the assumption of a logic-based model-theoretical description leads to \textit{``many non-trivial simplifications"} \cite{bao_towards_2011}.

In \cite{guler_semantic_2018}, the authors follow a different approach in the context of Natural Language Processing (NLP) defining semantic similarity %
as a semantic error measure to quantify the distance between the meanings of two words. Based on this metric, communication of a finite set of words is modeled as a Bayesian game from game theory and optimized for improved semantic transmission over a binary symmetric channel.

Recently, drawing inspiration from Weaver, Bao, Basu et al. \cite{weaver_recent_1949, bao_towards_2011, basu_preserving_2014} and enabled by the rise of ML in communications research, Deep Neural Network (DNN) - based NLP techniques, i.e., transformer networks, were introduced in Auto Encoders (AE) for the task of text and speech transmission \cite{farsad_deep_2018, xie_deep_2020, xie_deep_2021, weng_semantic_2021}. The aim of these techniques is to learn compressed hidden representations of the semantic content of sentences to improve communication efficiency, but exact recovery of the source (text) is the main objective. The approach improves performance in semantic metrics, especially at low SNR compared to classical digital transmissions.

Not considering Weaver's idea of semantic communication in particular, the authors in \cite{shao_learning_2022} show for the first time that task-oriented communications (Level C) for edge cloud transmission can be mathematically formulated as an Information Bottleneck (IB) optimization problem. Moreover, for solving the IB problem, they introduce a DNN-based approximation and show its applicability for the specific task of edge cloud transmission. The terminus \textit{``semantic information"} is only mentioned once in \cite{shao_learning_2022} referring to Joint Source Channel Coding (JSCC) of text from \cite{farsad_deep_2018} using recurrent neural networks.
In \cite{farsad_deep_2018}, the authors observe that sentences that express the same idea have embeddings that are close together in Hamming distance. But they use cross entropy between words and estimated words as the loss function and use the word error rate as the performance measure, which both do not reflect if two sentences have the same meaning but rather that both are exactly the same. %

As a result, semantic communication is still a nascent field: It remains still unclear what this term exactly means and especially its distinction from Joint Source Channel Coding (JSCC) \cite{farsad_deep_2018, bourtsoulatze_deep_2019}. As a result, many survey papers aim to provide an interpretation, see, e.g., \cite{popovski_semantic-effectiveness_2020, calvanese_strinati_6g_2021, lan_what_2021, uysal_semantic_2022, gunduz_beyond_2023}. We will revisit this issue in Sec.~\ref{sec:2}.

\subsection{Main Contributions}

The main contributions of this article are:
\begin{itemize}
	\item Motivated by the approach of Bao, Basu et al. \cite{bao_towards_2011, basu_preserving_2014}, we adopt the terminus of a semantic source. Inspired by Weaver's notion, we bring it to the context of communications by considering the complete Markov chain including semantic source, communications source, transmit signal, communication channel, and received signal in contrast to both \cite{bao_towards_2011, basu_preserving_2014}. Further, we also extend beyond the example of deterministic entailment relations between ``models" and ``messages" based on propositional logic in \cite{bao_towards_2011, basu_preserving_2014} to probabilistic semantic channels.
	\item We define the task of semantic communication in the sense that we perform data compression, coding, and transmission of messages observed such that the semantic Random Variable (\rv) at a recipient is best preserved. Basically, we implement joint source-channel coding of messages conveying the semantic \rv, but not differentiating between Levels A and B. We formulate the semantic communication design either as an Information Maximization or as an Information Bottleneck (IB) optimization problem \cite{tishby_information_2000, hassanpour_forward-aware_2020, hassanpour_forward-aware_2021}.
	\begin{itemize}
		\item Although the approach pursued here again leads to an IB problem as in \cite{shao_learning_2022}, our article introduces a new classification and perspective of semantic communication and different ML-based solution approaches. Different from \cite{shao_learning_2022}, we solve the IB problem maximizing the mutual information for a fixed encoder output dimension that bounds the information rate. %
		\item The publication presented here differs also both in the interpretation of what is meant by semantic information and in the objective of recovering this semantic information from approaches to semantic communication presented in the literature like, e.g., \cite{xie_deep_2021, beck_swarm_2023}.
	\end{itemize}
	\item Finally, we propose the ML-based semantic communication system \sinfoni\ for a distributed multipoint scenario in contrast to \cite{shao_learning_2022}: \sinfoni\ communicates the meaning behind multiple messages that are observed at different senders to a single receiver for semantic recovery. Compared to the distributed scenario in \cite{aguerri_distributed_2021, shao_task-oriented_2023}, we include the communication channel.
	\item We analyze \sinfoni\ by processing images as an example of messages. Notably, numerical results reveal a tremendous rate-normalized SNR shift up to $20$ dB compared to classically designed communication systems.
\end{itemize}
In the following, we reinterpret Weaver's philosophical considerations in Sec.~\ref{sec:21} paving the way for our proposed theoretical framework in Sec.~\ref{sec:2}. %
Finally, in Sec.~\ref{sec:3} and \ref{sec:conclusion}, we provide one numerical example of semantic communication, i.e., \sinfoni, and summarize the main results, respectively.

\section{A Framework for Semantics}
\label{sec:2}

\subsection{Philosophical Considerations}
\label{sec:21}

Despite the much-renewed interest, research on semantic communication is still in its infancy and recent work reveals a differing understanding of the word \emph{semantics}. In this work, we contribute our interpretation. To motivate it, we shortly revisit the research birth hour of communications from a philosophical point of view: Its theoretical foundation was laid by Shannon in his landmark paper \cite{shannon_mathematical_1948} in 1948.

He stated that \textit{``Frequently the messages have meaning; that is they refer to or are correlated according to some system with certain physical or conceptual entities. These semantic aspects of communication are irrelevant to the engineering problem."} In fact, this viewpoint abstracts all kinds of information one may transmit, e.g., oral and written speech, sensor data, etc., and lays also the foundation for the research area of Shannon information theory. Thus, it found its way into many other research areas where data or information is processed, including Artificial Intelligence (AI) and especially its subdomain Machine Learning (ML).

Weaver saw this broad applicability of Shannon's theory back in 1949. In his comprehensible review of \cite{shannon_mathematical_1948}, he first states that \textit{``there \textbf{seem} to be [communication] problems at three levels"} \cite{weaver_recent_1949} already mentioned in Sec.~\ref{sec:1_1}.
These three levels are quoted in recent works, where Level C is oftentimes referred to as goal-oriented communication instead \cite{calvanese_strinati_6g_2021}. %

But we note that in his concluding section, he then questions this segmentation: He argues for the generality of the theory at Level A for all levels and \textit{\textbf{``that the interrelation of the three levels is so considerable that one's final conclusion may be that the separation into the three levels is really artificial and undesirable"}}.

It is important to emphasize that the separation is rather arbitrary. We agree with Weaver's statement because the most important point that is also the focus herein is the definition of the term semantics, e.g., by Basu et al. \cite{bao_towards_2011, basu_preserving_2014}. Note that the entropy of the semantics is less than or equal to the entropy of the messages. Consequently, we can save information rate by introducing meaning or context. In fact, we are able to add arbitrarily many levels of semantic details to the communication problem and optimize communications for a specific semantic background, e.g., an application or human.

\subsection{Semantic System Model}
\label{sec:23}

\begin{figure*}[!t]%
	\centerline{\begin{tikzpicture}[%
					inner xsep = 0pt, inner ysep = 0pt, %
					outer xsep = 0pt, outer ysep = 0pt, %
					> = latex,
					ucircle/.style = {circle, line width = 0.30mm, minimum size = 4.00mm, draw = black, fill = white},
					urec/.style = {rectangle, line width = 0.60mm, draw = black, fill = white, minimum width = 50.0, minimum height = 20.00},
					invrec/.style = {rectangle, line width = 0.30mm, draw = none, fill = none, minimum width = 50.0, minimum height = 20.00},				
					arrowthick/.style = {->, line width = 0.6mm},
					arrowthin/.style = {<->, line width = 0.3mm},
					font = {\fontsize{7pt}{12}\selectfont},
					linesplit/.style = {circle, draw = black, fill = black, inner sep = 0pt, minimum height = 2},
					linethick/.style = {-, line width = 0.60mm},
					linethin/.style={-,line width = 0.30mm},
					]

\newcommand{\clsem}{blue} 					%
\newcommand{\clenc}{black} 					%
\newcommand{\clchan}{gray} 					%
\newcommand{\clencchan}{darkgreen} 			%
\newcommand{\cldec}{red} 						%

\node[urec, minimum height=40.00, minimum width=40.0, text width=2cm, align=center, rounded corners=5, draw = \clsem] (App1) at (-14,0) {Semantic Source\\ $\relvec\sim \prob(\relvec)$};
\node[urec, minimum height=40.00, minimum width=40.0, text width=2cm, align=center, rounded corners=5, draw = \clsem] (SChan) at (-11,0) {Semantic Channel\\ $\prob(\semvec|\relvec)$};
\node[urec, minimum height=30.00, minimum width=85.0, text width=3cm, align=center, rounded corners=5, draw = \clenc] (Enc) at (-7.5,0) {Encoder\\ $\prob_{\txpars}(\xvec|\semvec)=\dirac(\xvec-\encdet)$};
\node[urec, minimum height=30.00, minimum width=60.0, text width=2.2cm, align=center, rounded corners=5, draw = \clchan] (Chan) at (-4.5,-1.5) {Communication Channel $\prob(\yvec|\xvec)$};
\node[urec, minimum height=30.00, minimum width=171.0, text width=3cm, align=center, rounded corners=5, draw = \cldec] (Dec) at (-9,-5) {Semantic Decoder\\ $\aprob_{\rxpars}(\relvec|\yvec)$};
\node[urec, minimum height=30.00, minimum width=85.0, text width=2.25cm, align=center, rounded corners=5, draw = \clenc] (Dec2) at (-7.5,-3) {Classic Decoder\\ $\prob_{\txpars}(\semvec|\yvec)$};%
\node[urec, minimum height=40.00, minimum width=40.0, text width=2cm, align=center, rounded corners=5, draw = \clsem] (Inf) at (-11,-3) {Interpretation\\ $\prob(\relvec|\semvec)$};
\node[urec, minimum height=40.00, minimum width=40.0, text width=2cm, align=center, rounded corners=5, draw = \clsem] (App1_rx) at (-14,-3) {Semantic Estimate $\relvecest$};

\node[urec, text depth = 3 cm, minimum height=100.00, minimum width=190.0, fill=none, dashed, dash pattern={on 7pt off 3pt}, rounded corners=10, draw = \clencchan] (Enc2) at (-6.5,-0.5) {};
\node[anchor = north west, inner xsep = 0.00cm, inner ysep = 0.15cm, xshift = 1.5cm, text width=2cm, align=center] at (Enc2.north) {\color{\clencchan}{$\prob_{\txpars}(\yvec|\semvec)$}};

\draw[arrowthick] (App1) -> (SChan) node[midway, above, rotate=0, align=center, anchor=south, yshift = 0.1cm] {$\relvec$};
\draw[arrowthick] (SChan) -> (Enc) node[midway, above, rotate=0, align=center, anchor=south, yshift = 0.1cm] {$\semvec$};
\draw[arrowthick] (Enc) -| (Chan) node[midway, above, rotate=0, align=center, anchor=south, yshift = 0.1cm] {$\xvec$};
\node[linesplit, minimum height=5] (hnode) at (-4.5,-3) {};
\draw[linethick] (Chan) -> (hnode)  node[midway, above, rotate=0, align=center, anchor=west, xshift = 0.1cm] {$\yvec$};
\draw[arrowthick] (hnode) -> (Dec2) ;
\draw[arrowthick] (hnode) |- (Dec);
\draw[arrowthick] (Inf) -> (App1_rx) node[midway, above, rotate=0, align=center, anchor=south, yshift = 0.1cm] {$\relvec$};
\draw[arrowthick] (Dec2) -> (Inf) node[pos=0.5, above, rotate=0, yshift = 0.1cm] {$\semvec$};
\draw[arrowthick] (Dec) -| (App1_rx) node[pos=0.25, above, rotate=0, yshift = 0.1cm] {$\relvec$};

\end{tikzpicture}%
}
	\caption{Block diagram of the considered semantic system model.}
	\label{fig1:com_system_small}
\end{figure*}
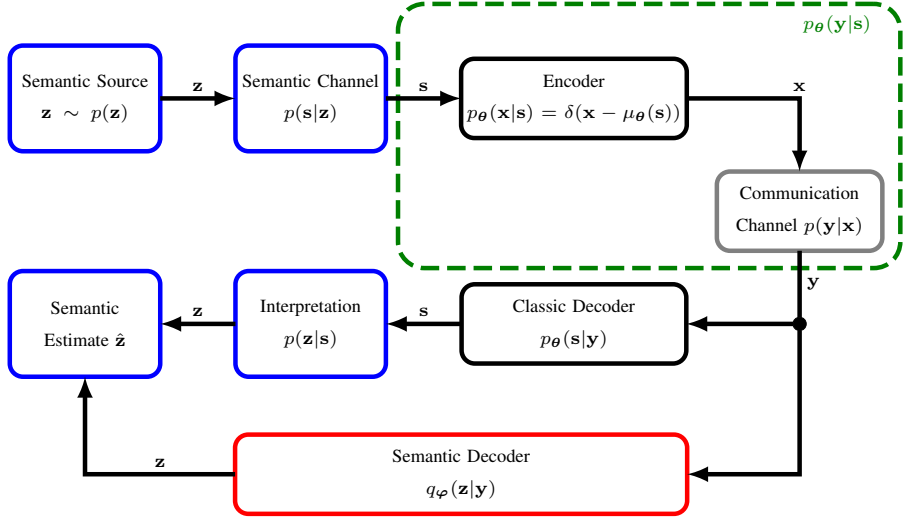

\subsubsection{Semantic Source and Channel}

Now, we will define our information-theoretic system model of semantic communication. Fig.~\ref{fig1:com_system_small} shows the schematic of our model. We assume the existence of a semantic source, described as a hidden target multivariate Random Variable (\rv) $\relvec\in\relset^{\Nrelvec\times 1}$ from a domain $\relset$ of dimension $\Nrelvec$ distributed according to a probability density or mass function (pdf/pmf) $\prob(\relvec)$. To simplify the discussion, we assume it to be discrete and memoryless.\footnote{For the remainder of the article, note that the domain of all \rv s $\dom$ may be either discrete or continuous. Further, we note that the definition of entropy for discrete and continuous \rv s differs. For example, the differential entropy of continuous \rv s may be negative whereas the entropy of discrete \rv s is always positive \cite{simeone2018brief}. Without loss of generality, we will thus assume all \rv s either to be discrete or to be continuous. In this work, we avoid notational clutter by using the expected value operator: Replacing the integral by summation over discrete \rv s, the equations are also valid for discrete \rv s and vice versa.}

This approach is similar to that of \cite{bao_towards_2011, basu_preserving_2014}: In \cite{bao_towards_2011, basu_preserving_2014}, the semantic source\footnote{Note that in \cite{basu_preserving_2014} the semantic information source is defined as a tuple $(\relvec,\semvec,\prob(\relvec,\semvec),\formlang)$. In this original notation, $\relvec$ is the model, $\semvec$ the message, $\prob(\relvec,\semvec)$ the joint distribution of $\relvec$ and $\semvec$, and $\formlang$ is the deterministic formal language.} is described by ``models of the world". In \cite{basu_preserving_2014}, a semantic channel then generates messages through entailment relations between ``models" and ``messages". We will call these ``messages" source signal and define it to be a \rv\ $\semvec\in\semset^{\Nsemvec\times 1}$ as it is usually observed and enters the communication system. In classic Shannon design, the aim is to reconstruct the source $\semvec$ as accurately as possible at the receiver side. Further, we note that the authors in \cite{basu_preserving_2014} considered the example of a semantic channel with deterministic entailment relations between $\relvec$ and $\semvec$ based on propositional logic. In this article, we go beyond this assumption and consider probabilistic semantic channels modeled by distribution $\prob(\semvec|\relvec)$ that include the entailment in \cite{basu_preserving_2014} as special cases, i.e., $\prob(\semvec|\relvec)=\dirac(\semvec-\funcrtrick(\relvec))$ where $\dirac(\cdot)$ is the Dirac delta function and $\funcrtrick(\cdot)$ is any generic function. Our viewpoint is motivated by the recent success of pattern recognition tools that advanced the field of AI in the $2010$s and may be used to extract semantics \cite{simeone2018very}.

Our approach also extends models as in \cite{xie_deep_2021}. There, the authors design a semantic communication system for the transmission of written language/text similar to \cite{farsad_deep_2018} using transformer networks. In contrast to our work, \cite{xie_deep_2021} does not define meaning as \rv\ $\relvec$. The objective in \cite{xie_deep_2021} is to reconstruct $\semvec$ (sentences) as well as possible, rather than the meaning (\rv\ $\relvec$) conveyed in $\semvec$. %
Optimization is done w.r.t. to a loss function consisting of two parts: Cross entropy between language input $\semvec$ and output estimate $\semhvec$, as well as a scaled mutual information term between transmit signal $\xvec$ and receive signal $\yvec$. After optimization, the authors measure semantic performance by some semantic metric $\Lossf(\semvec,\semhvec)$.

We now provide an example to explain what we understand under a semantic source $\relvec$ and channel $\prob(\semvec|\relvec)$: Let us assume a biologist who has an image of a tree. The biologist wants to know what kind of tree it is by interpreting the observed data (image). In this case, the semantic source $\relvec$ is a multivariate \rv\ composed of a categorical \rv\ with $\Nclass$ tree classes. For any realization (sample value) $\relvecreal_{\idxi}$ of the semantic source, the semantic channel $\prob(\semvec|\relvec)$ then outputs with some probability one image $\semvecreal_{\idxi}$ of a tree conveying characteristics of $\relvec$, i.e., its meaning. Note that the underlying meaning of the same sensed data (message) can be different for other recipients, e.g., humans or tasks/applications, i.e., in other semantic contexts. Imagine a child, i.e., a person with different characteristics (personality, expertise, knowledge, goals, and intentions) than the biologist, who is only interested if he/she can climb up this tree or whether the tree provides shade. Thus, we include the characteristics of the sender and receiver in \rv\ $\relvec$ and consider it directly in compression and encoding.

Compared to \cite{bao_towards_2011}, we, therefore, argue that we also include level C by semantic source and channel since context can be included on increasing layers of complexity. First, a \rv\ $\relvec_1$ might capture the interpretation, like the classification of images or sensor data. Moving beyond the first semantic layer, then a \rv\ $\relvec_2$ might expand this towards a more general goal, like keeping a constant temperature in power plant control. In fact, we can add or remove context, i.e., semantics and goals, arbitrarily often according to the human or application behind, and we can optimize the overall (communication) system w.r.t. $\relvec_1,\relvec_2,\dots,\relvec_i$, respectively.

As a last remark, we note that we basically defined probabilistic semantic relationships, and it remains the question of how exactly they might look. In our example, the meaning of the images needs to be labeled into real-world data pairs $\{\semvecreal_{\idxi},\relvecreal_{\idxi}\}$ by experts/humans, since image recognition lacks precise mathematical models. This is also true for NLP \cite{xie_deep_2021}: How can we measure if two sentences have the same meaning, i.e., how does the semantic space look like?
In contrast, in \cite{basu_preserving_2014}, the authors are able to solve their well-defined technical problem (motion detection) by a model-driven approach. We can thus distinguish between model and data-driven semantics, which both can be handled within Shannon's information theory.

\subsubsection{Semantic Channel Encoding}

After the semantic source and channel in Fig.~\ref{fig1:com_system_small}, we extend upon \cite{bao_towards_2011} by differentiating between ``message"/source signal $\semvec$ and transmit signal $\xvec\in\xset^{\ntx\times 1}$. Our challenge is to encode the source signal $\semvec$ onto the transmit signal vector $\xvec$ for reliable semantic communication through the physical communication channel $\prob(\yvec|\xvec)$, where $\yvec\in\yset^{\Nrx\times 1}$ is the received signal vector. We assume the encoder $\prob_{\txpars}(\xvec|\semvec)$ to be parametrized by a parameter vector $\txpars\in\realnum^{\Ntxpars\times 1}$. Note that $\prob_{\txpars}(\xvec|\semvec)$ is probabilistic here, but assumed to be deterministic in communications with $\prob_{\txpars}(\xvec|\semvec)=\dirac(\xvec-\encdet)$ and encoder function $\encdet$. %

In summary, in contrast to both \cite{bao_towards_2011, basu_preserving_2014}, we consider the complete Markov chain $\relvec \leftrightarrow \semvec \leftrightarrow \xvec \leftrightarrow \yvec$ including semantic source $\relvec$, communications source $\semvec$, transmit signal $\xvec$ and receive signal $\yvec$. By this means, we distinguish from \cite{basu_preserving_2014} which only deals with semantic compression, and \cite{bao_towards_2011} which is about joint semantic compression and channel coding (Level B). In \cite{bao_towards_2011}, the authors consider the classic transmission system (Level A) as the (semantic) channel (not to be confused with the definition of the semantic channel in \cite{basu_preserving_2014} which we make use of in this publication).

At the receiver side, one approach is maximum a posteriori decoding w.r.t. \rv\ $\semvec$ that uses the posterior $\prob_{\txpars}(\semvec|\yvec)$, being deduced from prior $\prob(\semvec)$ and likelihood $\prob_{\txpars}(\yvec|\semvec)$ by application of Bayes law. Based on the estimate of $\semvec$, then the receiver interprets the actual semantic content $\relvec$ by $\prob(\relvec|\semvec)$.

Another approach we propose is to include the semantic hidden target \rv\ $\relvec$ into the design by processing $\prob_{\txpars}(\relvec|\yvec)$. If the calculation of the posterior is intractable, we can replace $\prob_{\txpars}(\relvec|\yvec)$ by the approximation $\aprob_{\rxpars}(\relvec|\yvec)$, i.e., the semantic decoder, with parameters $\rxpars\in\realnum^{\Nrxpars\times 1}$. We expect the following benefit: We assume the entropy $\HE[\relvec]=\evalnb[\relvec\sim\prob(\relvec)]{-\ln \prob(\relvec)}$ of the semantic \rv\ $\relvec$, i.e., the actual semantic uncertainty or information content, to be less or equal to the entropy $\HE[\semvec]$ of the source $\semvec$, i.e., $\HE[\relvec]\le\HE[\semvec]$. There, $\eval[\xvec\sim \prob(\xvec)]{\funcrtrick(\xvec)}$ denotes the expected value of $\funcrtrick(\xvec)$ w.r.t. both discrete or continuous \rv s $\xvec$. Consequently, since we would like to preserve the relevant, i.e., semantic, \rv\ $\relvec$ rather than $\semvec$, we can compress more s.t. preserving $\relvec$ conveyed in $\semvec$. Note that in semantic communication the relevant variable is $\relvec$, not $\semvec$. Thus, processing $\prob_{\txpars}(\semvec|\yvec)$ without taking $\relvec$ into consideration resembles the classical approach. Instead of using (and transmitting) $\semvec$ for inference of $\relvec$, we now want to find a compressed representation $\yvec$ of $\semvec$ containing the relevant information about $\relvec$.

\subsection{Semantic Communication Design via Infomax Principle}
\label{sec:infomaxlearning}

After explaining the system model and the basic components, we are able to approach a semantic communication system design: %
We first define an optimization problem to obtain the encoder $\prob_{\txpars}(\xvec|\semvec)$ following the Infomax principle from an information theoretic perspective \cite{simeone2018brief}. Thus, we like to find the distribution $\prob_{\txpars}(\xvec|\semvec)$ that maps $\semvec$ to a representation $\xvec$ such that most information of the relevant \rv\ $\relvec$ is included in $\yvec$, i.e., we maximize the Mutual Information (MI) $\mi{\relvec;\yvec}$ w.r.t. $\prob_{\txpars}(\xvec|\semvec)$ \cite{vincent_stacked_2010}:
\begin{align}
	&\argmax[\prob_{\txpars}(\xvec|\semvec)]\mi[\txpars]{\relvec;\yvec} \label{eq:infomax} \\
	=&\argmax[\prob_{\txpars}(\xvec|\semvec)]\eval[\relvec,\yvec\sim\prob_{\txpars}(\relvec,\yvec)]{\ln \frac{\prob_{\txpars}(\relvec,\yvec)}{\prob(\relvec)\prob_{\txpars}(\yvec)}} \\
	=&\argmax[\prob_{\txpars}(\xvec|\semvec)]\HE[\relvec] - \HE[\prob_{\txpars}(\relvec,\yvec), \prob_{\txpars}(\relvec|\yvec)] \label{eq:infomax_connection} \\
	=&\argmax[\prob_{\txpars}(\xvec|\semvec)]\eval[\relvec,\yvec\sim\prob_{\txpars}(\relvec,\yvec)]{\ln \prob_{\txpars}(\relvec|\yvec)} \eqpoint \label{eq:infomax2} %
\end{align}
There, $\HE[\prob(\xvec), \aprob(\xvec)]=\evalnb[\xvec\sim\prob(\xvec)]{-\ln \aprob(\xvec)}$ is the cross entropy between two pdfs/pmfs $\prob(\xvec)$ and $\aprob(\xvec)$. Note independence from $\txpars$ in $\HE[\relvec]$ and dependence in $\prob_{\txpars}(\relvec|\yvec)$ and $\prob_{\txpars}(\relvec,\yvec)$ through the Markov chain $\relvec \rightarrow \semvec \rightarrow \yvec$. It is worth mentioning that we so far have not set any constraint on the variables we deal with. Hence, the form of $\prob_{\txpars}(\yvec|\semvec)$ has to be constrained to avoid learning a trivial identity mapping $\yvec=\semvec$. We indeed constrain the optimization by our communication channel $\prob(\yvec|\xvec)$ we assume to be given.

If the calculation of the posterior $\prob_{\txpars}(\relvec|\yvec)$ in \eqref{eq:infomax2} is intractable, we are able to replace it by a variational distribution $\aprob_{\rxpars}(\relvec|\yvec)$ with parameters $\rxpars$. Similar to the transmitter, DNNs are usually proposed \cite{xie_deep_2021, sana_learning_2022} for the design of the approximate posterior $\aprob_{\rxpars}(\relvec|\yvec)$ at the receiver. To improve the performance complexity trade-off, the application of \textit{deep unfolding} can be considered, a model-driven learning approach that introduces model knowledge of $\prob_{\txpars}(\semvec,\xvec,\yvec,\relvec)$ to create $\aprob_{\rxpars}(\relvec|\yvec)$ \cite{beck_cmdnet_2021, farsad_data-driven_2021}.
With $\aprob_{\rxpars}(\relvec|\yvec)$, we are able to define a Mutual Information Lower BOund (MILBO) \cite{vincent_stacked_2010} similar to the well-known Evidence Lower BOund (ELBO) \cite{simeone2018very}:
\begin{align}
	\mi[\txpars]{\relvec;\yvec}&\geq \eval[\relvec,\yvec\sim\prob_{\txpars}(\relvec,\yvec)]{\ln \aprob_{\rxpars}(\relvec|\yvec)} \label{eq:milbo} \\
	&=\eval[\yvec\sim\prob_{\txpars}(\yvec)]{\eval[\relvec\sim\prob_{\txpars}(\relvec|\yvec)]{\ln \aprob_{\rxpars}(\relvec|\yvec)}} \\
	&= -\eval[\yvec\sim\prob_{\txpars}(\yvec)]{\HE[\prob_{\txpars}(\relvec|\yvec), \aprob_{\rxpars}(\relvec|\yvec)]} \label{eq:ce_loss} \\
	&= -\ceparams  \eqpoint \label{eq:ce_loss_opt}
\end{align}
The lower bound holds since $- \HE[\prob_{\txpars}(\relvec,\yvec), \prob_{\txpars}(\relvec|\yvec)]$ itself is a lower bound of the expression in \eqref{eq:infomax_connection} and $\eval[\relvec,\yvec\sim\prob_{\txpars}(\relvec,\yvec)]{\ln \prob_{\txpars}(\relvec|\yvec)/\ln \aprob_{\rxpars}(\relvec|\yvec)}\geq 0$. Now, we can calculate optimal values of $\txpars$ and $\rxpars$ of our semantic communication design by minimizing the amortized cross entropy $\ceparams$ in \eqref{eq:ce_loss}, i.e., marginalized across observations $\yvec$ \cite{beck_cmdnet_2021}.

Thus, the idea is to learn parametrizations of the transmitter discriminative model and of the variational receiver posterior, e.g., by Auto Encoders (AEs) or reinforcement learning. Note that in our semantic problem \eqref{eq:infomax}, we do not auto encode the hidden $\relvec$ itself, but encode $\semvec$ to obtain $\relvec$ by decoding. This can be seen from Fig.~\ref{fig1:com_system_small} and by rewriting the amortized cross entropy \eqref{eq:ce_loss}, \eqref{eq:ce_loss_opt}:
\begin{align}
	\ceparams&=\eval[\yvec\sim\prob(\yvec)]{\HE[\prob_{\txpars}(\relvec|\yvec), \aprob_{\rxpars}(\relvec|\yvec)]} \label{eq:ce_amortized} \\
	&= \eval[\semvec,\xvec,\yvec,\relvec\sim \prob_{\txpars}(\semvec,\xvec,\yvec,\relvec)]{-\ln \aprob_{\rxpars}(\relvec|\yvec)} \label{eq:relvar_infomax}\\
	&= \eval[\semvec,\relvec\sim\prob(\semvec,\relvec)]{\eval[\xvec\sim\prob_{\txpars}(\xvec|\semvec)]{\eval[\yvec\sim\prob(\yvec|\xvec)]{-\ln \aprob_{\rxpars}(\relvec|\yvec)}}} \eqpoint \nonumber
\end{align}
We can further prove the amortized cross entropy to be decomposable into%
\begin{align}
	&\ceparams  \nonumber \\ %
	=&\E_{\yvec\sim\prob_{\txpars}(\yvec)}\big[\E_{\relvec\sim\prob_{\txpars}(\relvec|\yvec)}[-\ln \aprob_{\rxpars}(\relvec|\yvec)+\ln \prob_{\txpars}(\relvec|\yvec) \nonumber \\
	&-\ln \prob_{\txpars}(\relvec|\yvec)]\big] \\
	=&\eval[\yvec\sim\prob_{\txpars}(\yvec)]{\dkl{\prob_{\txpars}(\relvec|\yvec)}{\aprob_{\rxpars}(\relvec|\yvec)}} + \underbrace{\HE[\relvec|\yvec]}_{=-\mi[\txpars]{\relvec;\yvec} + \HE[\relvec]} \\
	=&\HE[\relvec] - \underbrace{\mi[\txpars]{\relvec;\yvec}}_{\txt{enc. objective}} + \underbrace{\eval[\yvec\sim\prob_{\txpars}(\yvec)]{\dkl{\prob_{\txpars}(\relvec|\yvec)}{\aprob_{\rxpars}(\relvec|\yvec)}}}_{\txt{dec. objective}}. \label{eq:aeloss} %
\end{align}
To the end, maximization of the MILBO w.r.t. $\txpars$ and $\rxpars$ balances maximization of the mutual information $\mi[\txpars]{\relvec;\yvec}$ and minimization of the Kullback-Leibler (KL) divergence $\dkl{\prob_{\txpars}(\relvec|\yvec)}{\aprob_{\rxpars}(\relvec|\yvec)}$. The former objective can be seen as a regularization term that favors encoders with high mutual information, for which decoders can be learned that are close to the true posterior.

\subsection{Classical Design Approach}

If we consider classical communication design approaches, we would solve the problem
\begin{align}
	&\argmax[\prob_{\txpars}(\xvec|\semvec)]\mi{\semvec;\yvec} \label{eq:jscc}
\end{align}
which relates to Joint Source Channel Coding (JSCC). There, the aim is to find a representation $\xvec$ that retains a significant amount of information about the source signal $\semvec$ in $\yvec$. Again, we can apply the lower bound \eqref{eq:ce_loss_opt}. In fact, bounding Eq. \eqref{eq:jscc} by \eqref{eq:ce_loss_opt} shows that approximate maximization of the mutual information justifies the minimization of the cross entropy in the Auto Encoder (AE) approach \cite{oshea_introduction_2017} oftentimes seen in recent wireless communication literature \cite{oshea_introduction_2017, bourtsoulatze_deep_2019, farsad_deep_2018}. %

\subsection{Information Bottleneck View}

It should be stressed that we have not set any constraints on the variables in the Infomax problem so far. However, in many applications, compression is needed because of the limited information rate. Therefore, we can formulate an optimization problem where we like to maximize the relevant information $\mi[\txpars]{\relvec;\yvec}$ subject to the constraint to limit the compression rate $\mi[\txpars]{\semvec;\yvec}$ to a maximum information rate $\mic$:
\begin{align}
	&\argmax[\prob_{\txpars}(\xvec|\semvec)]\mi[\txpars]{\relvec;\yvec} \st \mi[\txpars]{\semvec;\yvec}\le \mic  \eqpoint \label{eq:ibm}
\end{align}
Problem \eqref{eq:ibm} is called the Information Bottleneck (IB) problem \cite{calvanese_strinati_6g_2021, tishby_information_2000, goldfeld_information_2020, zaidi_information_2020}. %
The IB method introduced by Tishby et al. \cite{tishby_information_2000} has been the subject of intensive research for years and has proven to be a suitable mathematical/information-theoretical framework for solving numerous problems - also in wireless communications \cite{kurkoski_quantization_2014, lewandowsky_ldpc_2018, hassanpour_forward-aware_2020, hassanpour_forward-aware_2021}. Note that we aim for an encoder that compresses $\semvec$ into a compact representation $\xvec$ for discrete \rv s by clustering and for continuous \rv s by dimensionality reduction.

To solve the constrained optimization problem \eqref{eq:ibm}, we can use Lagrangian optimization and obtain
\begin{align}
	&\argmax[\prob_{\txpars}(\xvec|\semvec)]\mi[\txpars]{\relvec;\yvec}-\lagrange \mi[\txpars]{\semvec;\yvec} \label{eq:ibm2}
\end{align}
with Lagrange multiplier $\lagrange\ge 0$. The Lagrange multiplier $\lagrange$ allows defining a trade-off between the relevant information $\mi[\txpars]{\relvec;\yvec}$ and compression rate $\mi[\txpars]{\semvec;\yvec}$, which indicates the relation to rate distortion theory \cite{hassanpour_forward-aware_2020}. With $\lagrange=0$, we have the InfoMax problem \eqref{eq:infomax} whereas for $\lagrange\to\infty$ we minimize compression rate. Calculation of the mutual information terms may be computationally intractable, as in the InfoMax problem \eqref{eq:infomax}. Approximation approaches can be found in \cite{alemi_deep_2019, belghazi_mine_2018}. Notable exceptions include if the \rv s are all discrete or Gaussian distributed.

We note that in \cite{calvanese_strinati_6g_2021, shao_learning_2022} the authors already introduced the IB problem to task-oriented communications. But \cite{calvanese_strinati_6g_2021, shao_learning_2022} do not address our viewpoint or classification: We compress and channel encode the messages/communications source $\semvec$ for given entailment $\prob(\semvec|\relvec)$, in the sense of a data-reduced and reliable communication of the semantic \rv\ $\relvec$. Basically, we implement joint source-channel coding of $\semvec$ s.t. preserving the semantic \rv\ $\relvec$, and we do not differentiate between Levels A and B as indicated by Weaver's notion outlined in Sec.~\ref{sec:1_1}. Indeed, we draw a direct connection to IB compared to related semantic communication literature \cite{farsad_deep_2018, xie_deep_2021, sana_learning_2022} that so far only included optimization with terms reminiscent of the IB problem.

This article does not only distinct itself on a conceptual, but also on a technical level from \cite{shao_learning_2022, shao_task-oriented_2023}: We follow a different strategy to solve \eqref{eq:ibm}.

First, using the data processing inequality \cite{cover_elements_2006}, we see that the compression rate is upper bounded by the mutual information of the encoder $\mi[\txpars]{\semvec;\xvec}$ and that of the channel $\mi{\xvec;\yvec}$:
\begin{align}
	\mi[\txpars]{\semvec;\yvec}\le \min \lcb \mi[\txpars]{\semvec;\xvec}, \mi{\xvec;\yvec} \rcb \eqpoint
\end{align}
In case of negligible encoder compression $\mi[\txpars]{\semvec;\xvec}>\mi{\xvec;\yvec}$, the channel becomes the limiting factor of information rate. For example, with a deterministic continuous mapping $\xvec=\encdet$, this is true since $\mi[\txpars]{\semvec;\xvec}\rightarrow\infty$. Using the chain rule of mutual information \cite{cover_elements_2006}, we see that this upper bound on compression rate grows with the dimension of $\xvec$, i.e., the number of channel uses $\ntx$:
\begin{align}
	\mi[\txpars]{\semvec;\yvec}\le \mi{\xvec;\yvec} = \summ[\ntx]{\indx=1} \underbrace{\mi{\xvar_{\indx};\yvec|\xvar_{\indx-1},\dots,\xvar_{1}}}_{\ge 0} \eqpoint \label{eq:mi_sum}
\end{align}
Assuming $\yvec$ to be conditional dependent on $\xvar_{\indx}$ given $\xvar_{\indx-1},\dots,\xvar_{1}$, i.e., $\prob(\yvec|\xvar_{\indx},\dots,\xvar_{1})\neq \prob(\yvec|\xvar_{\indx-1},\dots,\xvar_{1})$ being, e.g., true for an AWGN channel, it is $\mi{\xvar_{\indx};\yvec|\xvar_{\indx-1},\dots,\xvar_{1}}>0$ \cite{cover_elements_2006} and the sum in \eqref{eq:mi_sum} indeed strictly increases. Replacing $\yvec$ in $\mi{\xvec;\yvec}$ of \eqref{eq:mi_sum} by $\semvec$, the result also holds for encoder compression $\mi[\txpars]{\semvec;\xvec}$, respectively. Hence, increasing the encoder output dimension $\ntx$, we can increase the possible compression rate $\mi[\txpars]{\semvec;\yvec}$. Interchanging $\xvec$ and $\yvec$ in \eqref{eq:mi_sum}, we see that the same holds for the receiver input dimension $\Nrx$.

Further, the mutual information of the channel and thus the compression rate are upper bounded by channel capacity:
\begin{align}
	\mi[\txpars]{\semvec;\yvec}\le \mi{\xvec;\yvec} \le \max_{\prob(\xvec); \eval[]{\abs[\xvar_{\indx}]^2}\le 1} \mi{\xvec;\yvec}=\capa \eqpoint \label{eq:mi_channel_capacity} %
\end{align}
For example, with an AWGN channel, we have $\capa=\ntx/2\cdot \ln\lpa 1 + 1/\noisestd^2\rpa$ again increasing with $\ntx$.

Now, let us assume the \rv s to be discrete so that $\HE[\xvec|\semvec]\ge 0$. Indeed, this is true if the \rv s are processed discretely with finite resolution on digital signal processors, as in the numerical example of Sec.~\ref{sec:3}. As long as $\mi[\txpars]{\semvec;\xvec}<\capa$, all information of the discrete \rv s can be transmitted through the channel with arbitrary low error probability according to Shannon's channel coding theorem \cite{shannon_mathematical_1948}. Then, we can upper bound encoder compression $\mi[\txpars]{\semvec;\xvec}$ and thus compression rate $\mi[\txpars]{\semvec;\yvec}$ by the sum of entropies of any output $\xvar_{\indx}$ \cite{cover_elements_2006} of the encoder $\prob_{\txpars}(\xvec|\semvec)$ - each with cardinality $\abs[\xset]$:
\begin{align}
	\mi[\txpars]{\semvec;\xvec}= \HE[\xvec] - \underbrace{\HE[\xvec|\semvec]}_{\ge 0} \le \HE[\xvec]
	&\le \summ[\ntx]{\indx=1}\HE[\xvar_{\indx}] \\
	&\le\ntx\cdot \log_2(\abs[\xset]) \eqpoint \label{eq:entropy_independent}
\end{align}
Note that the entropy sum in \eqref{eq:entropy_independent} grows again with $\ntx$ for discrete \rv s since $0\le\HE[\xvar_{\indx}]\le \log_2(\abs[\xset])$. Moreover, we can define an encoder capacity $\capa_{\txpars}$ analogous to channel capacity $\capa$ in \eqref{eq:mi_channel_capacity} upper bounding encoder compression $\mi[\txpars]{\semvec;\xvec}$. It may be restricted by the chosen (DNN) model $\prob_{\txpars}(\xvec|\semvec)$ and optimization procedure w.r.t. $\txpars$, i.e., the hypothesis class \cite{simeone2018very}.

In summary, we have proven by \eqref{eq:mi_channel_capacity} and \eqref{eq:entropy_independent} that there is an information bottleneck when maximizing the relevant information $\mi[\txpars]{\relvec;\yvec}$ either due to the channel distortion $\mi{\xvec;\yvec}$ or encoder compression $\mi[\txpars]{\semvec;\yvec}$.

To fully exploit the available resources, we set constraint $\mic$ to be equal to the upper bound, i.e., channel capacity $\capa$ or the upper bound on encoder compression rate $\ntx\cdot \log_2(\abs[\xset])$. In both cases, the upper bound grows (linearly) with the encoder output dimension $\ntx$, and thus we can set the constraint $\mic$ higher or lower by choosing $\ntx$.

With fixed constraint $\mic$, we maximize the relevant information $\mi[\txpars]{\relvec;\yvec}$. By doing so, we derive an exact solution to \eqref{eq:ibm} that maximizes $\mi[\txpars]{\relvec;\yvec}$ for a fixed encoder output dimension that bounds the compression rate. As in the InfoMax problem, we can exploit the MILBO to use the amortized cross entropy $\ceparams$ in \eqref{eq:ce_amortized} as the optimization criterion.

In \cite{shao_learning_2022}, however, the authors solve the variational IB problem of \eqref{eq:ibm2} and require tuning of $\lagrange$. Albeit also using the MILBO as a variational approximation to the first term in \eqref{eq:ibm2}, they introduce a KL divergence term as an upper bound to compression rate $\mi[\txpars]{\semvec;\yvec}$ derived by $\dkl{\prob_{\txpars}(\yvec)}{\aprob_{\varpars}(\yvec)}\ge 0$ with some variational distribution $\aprob_{\varpars}(\yvec)$ with parameters $\varpars$ \cite{alemi_deep_2019}.
Then, the variational IB objective function reads \cite{alemi_deep_2019}:
\begin{align}
	&\mi[\txpars]{\relvec;\yvec}-\lagrange \mi[\txpars]{\semvec;\yvec} \nonumber \\
	\ge& \eval[\relvec,\yvec\sim\prob_{\txpars}(\relvec,\yvec)]{\ln \aprob_{\rxpars}(\relvec|\yvec)} \nonumber \\
	&- \lagrange \eval[\semvec\sim\prob(\semvec)]{\dkl{\prob_{\txpars}(\yvec|\semvec)}{\aprob_{\varpars}(\yvec)}} \label{eq:vibm} \eqpoint
\end{align}
Moreover, the authors use a log-uniform distribution as the variational prior $\aprob_{\varpars}(\yvec)$ in \cite{shao_learning_2022} to induce sparsity on $\yvec$ so that the number of outputs is dynamically determined based on the channel condition or SNR, i.e., $\prob_{\txpars}(\yvec|\semvec,\noisestd^2)$. The approach additionally necessitates approximation of the KL divergence term in \eqref{eq:vibm} and estimation of the noise variance $\noisestd^2$.

With our approach we avoid the additional approximations and tuning of the hyperparameter $\lagrange$ in \eqref{eq:vibm} possibly enabling better semantic performance as well as reduced inference and training complexity at the cost of full usage of $\ntx$ channels even when the channel capacity $\capa$ allows for its reduction. We leave a numerical comparison to \cite{shao_learning_2022} for future research as this is out of the scope of this paper.

\subsection{Implementation Considerations}
\label{sec:3impl}

Now, we will provide important implementation considerations for optimization of \eqref{eq:ce_loss_opt}/\eqref{eq:relvar_infomax} and \eqref{eq:ibm}. We note that computation of the MILBO leads to similar problems as for the ELBO \cite{simeone2018brief}: If calculating the expected value in \eqref{eq:relvar_infomax} cannot be solved analytically or is computationally intractable - as typically the case with DNNs -, we can approximate it using Monte Carlo sampling techniques with $\Nsamp$ samples $\lcb\relvecreal_{\indi},\semvecreal_{\indi},\xvecreal_{\indi},\yvecreal_{\indi}\rcb_{i=1}^{\Nsamp}$.

For Stochastic Gradient Descent (SGD) - based optimization like, e.g., in the AE approach, the gradient w.r.t. $\rxpars$ can then be calculated by
\begin{align}
	\frac{\partial\ceparams}{\partial\rxpars}=&\frac{\partial}{\partial\rxpars}\eval[\relvec,\semvec,\yvec\sim\prob_{\txpars}(\yvec|\semvec)\prob(\semvec|\relvec)\prob(\relvec)]{-\ln \aprob_{\rxpars}(\relvec|\yvec)} \\
	=&-\eval[\relvec,\semvec,\yvec\sim\prob_{\txpars}(\yvec|\semvec)\prob(\semvec|\relvec)\prob(\relvec)]{\frac{\partial \ln\aprob_{\rxpars}(\relvec|\yvec)}{\partial\rxpars}} \\
	\approx& -\frac{1}{\Nsamp}\summ[\Nsamp]{\indi=1}\frac{\partial\ln \aprob_{\rxpars}(\relvecreal_{\indi}|\yvecreal_{\indi})}{\partial \rxpars} \label{eq:sgdrxpars}
\end{align}
with $\Nsamp$ being equal to the batch size $\Nb$ and by application of the backpropagation algorithm to $\frac{\partial}{\partial \rxpars}\ln \aprob_{\rxpars}(\relvecreal_{\indi}|\yvecreal_{\indi})=\frac{\partial}{\partial \rxpars}\aprob_{\rxpars}(\relvecreal_{\indi}|\yvecreal_{\indi})/\aprob_{\rxpars}(\relvecreal_{\indi}|\yvecreal_{\indi})$
in Automatic Differentiation Frameworks (ADF), e.g., TensorFlow and PyTorch. Computation of the so-called \reinforce\ gradient w.r.t. $\txpars$ leads to a high variance of the gradient estimate since we sample w.r.t. the distribution $\prob_{\txpars}(\yvec|\semvec)$ dependent on $\txpars$ \cite{simeone2018brief}.

Leveraging the direct relationship between $\txpars$ and $\yvec$ in $\ln \aprob_{\rxpars}(\relvec|\yvec)$ can help reduce the estimator's high variance. Typically, e.g., in Variational AEs (VAE), the reparametrization trick is used to achieve this \cite{simeone2018brief}. Here we can apply it if we can decompose the latent variable $\yvec\sim\prob_{\txpars}(\yvec|\semvec)$ into a differentiable function $\yvec=\funcrtrick_{\txpars}(\semvec,\noisevec)$ and a \rv\ $\noisevec\sim \prob(\noisevec)$ independent of $\txpars$. Fortunately, the typical forward model of a communication system $\prob_{\txpars}(\yvec|\semvec)$ fulfills this criterion. Assuming a deterministic DNN encoder $\xvec=\encdet$ and additive noise $\noisevec$ with covariance $\covariance$, we can thus rewrite $\yvec$ into $\funcrtrick_{\txpars}(\semvec, \noisevec)=\encdet+\covariance^{1/2}\cdot \noisevec$ and accordingly the amortized cross entropy gradient into:
\begin{align}
	\frac{\partial\ceparams}{\partial\txpars} %
	=&-\frac{\partial}{\partial\txpars}\eval[\relvec,\semvec,\yvec\sim\prob_{\txpars}(\yvec|\semvec)\prob(\semvec,\relvec)]{\ln\aprob_{\rxpars}(\relvec|\yvec)} \\
	=&-\eval[\relvec,\semvec,\noisevec\sim\prob(\noisevec)\prob(\semvec|\relvec)\prob(\relvec)]{\frac{\partial \funcrtrick_{\txpars}(\semvec,\noisevec)}{\partial\txpars} \cdot \frac{\partial \ln \aprob_{\rxpars}(\relvec|\yvec)}{\partial\yvec}} \\
	\approx& -\frac{1}{\Nsamp}\summ[\Nsamp]{\indi=1}\frac{\partial \funcrtrick_{\txpars}(\semvecreal_{\indi},\noisevecreal_{\indi})}{\partial\txpars}\cdot \frac{\partial \ln \aprob_{\rxpars}(\relvecreal_{\indi}|\yvecreal_{\indi})}{\partial\yvec}\bigg|_{\yvec=\funcrtrick_{\txpars}(\semvecreal_{\indi}, \noisevecreal_{\indi})} \label{eq:reparam} \eqpoint
\end{align}
The trick can be easily implemented in ADFs by adding a noise layer after (DNN) function $\xvec=\encdet$, typically used for regularization in ML literature. Then, our loss function \eqref{eq:relvar_infomax} amounts to
\begin{align}
	\ceparams%
	\approx -\frac{1}{\Nsamp}\summ[\Nsamp]{\indi=1}\ln \aprob_{\rxpars}(\relvecreal_{\indi}|\yvec_{\indi}=\funcrtrick_{\txpars}(\semvecreal_{\indi}, \noisevecreal_{\indi})) \eqpoint \label{eq:mcloss}
\end{align}
This allows for joint optimization of both $\txpars$ and $\rxpars$, as demonstrated in recent works \cite{oshea_introduction_2017}, treating unsupervised optimization of AEs as a supervised learning problem.

\section{Example of Semantic Information Recovery}
\label{sec:3}

\begin{figure}[!t]
	\centerline{\begin{tikzpicture}[%
					inner xsep=0pt, inner ysep=0pt, %
					outer xsep=0pt, outer ysep=0pt, %
					>=latex,
					ucircle/.style={circle,line width=0.30mm,minimum size=4.00mm,draw=black,fill=white},
					urec/.style={rectangle,line width=0.60mm,draw=black,fill=white,minimum width=50.0,minimum height=20.00},
					invrec/.style={rectangle,line width=0.30mm,draw=none,fill=none,minimum width=50.0,minimum height=20.00},
					arrowthick/.style={->,line width=0.6mm},
					arrowthin/.style={<->,line width=0.6mm},
					linesplit/.style={circle, draw=black, fill=black, inner sep=0pt, minimum height=2},
					font={\fontsize{7pt}{12}\selectfont},
					]

\newcommand{\clsem}{blue} 				%
\newcommand{\clsemchan}{blue} 		%
\newcommand{\clenc}{black} 				%
\newcommand{\clresnet}{\clamp} 		%
\newcommand{\clchan}{gray} 				%
\newcommand{\cltrans}{purple} 			%
\newcommand{\cldec}{red} 					%

\node[urec, draw = \clsem, minimum height=10.00, minimum width=10.0, text width=1.3cm, align=center, rounded corners=5] (z0) at (-14.50, -4.25) {$\relvec$};

\node[urec, draw = \clsemchan, minimum height = 25.0, minimum width=40.0, text width=1.3cm, align=center, rounded corners=5] (image1) at (-14.5,-2.55) {Sem. Chan. $\prob(\semvec|\relvec)$};
\node[urec, draw=\clamp, minimum height=30.00, minimum width=50.0, text width=1.8cm, align=center, rounded corners=5] (resnet1) at (-10.125,-2.55) {ResNet Feature Extractor};
\draw[arrowthick] (image1) -> (resnet1) node[pos=0.42, rotate=0, align=left, anchor=south, xshift = 0.0cm, yshift = 0.1cm, rounded corners=5] {Image:\\$\semvec \in \realnum^{\Nimx\times \Nimy\times \Nimc}$};
\draw[arrowthick] (z0) -> (image1);

\node[urec, draw = \clresnet, minimum height=20.00, minimum width=60.0, text width=1.3cm, align=center, rounded corners=5] (cl) at (-10.125,-4.25) {Classifier};
\node[urec, draw = \clsem, minimum height=10.00, minimum width=10.0, text width=1.3cm, align=center, rounded corners=5] (z) at (-10.125,-5.5) {$\relvecest$};
\node[urec, draw = \clenc, minimum height=90.00, minimum width=80.0, text width=2.25cm, fill=none, dashed, dash pattern={on 7pt off 3pt}, rounded corners=5, align = left] (Dec) at (-10.125, -3.30) {};
\node[anchor = east, inner sep = 0.00cm, yshift = 0.0cm, xshift = 0.0cm, text width=2.00cm, align=center] at (Dec.west) {ResNet classifier:\\ $\aprob_{\rxpars}(\relvec|\semvec)$};

\draw[arrowthick] (resnet1) -> (cl) node[midway, rotate=0, align=center, anchor=west, xshift = 0.1cm, yshift = 0.1cm] {$\featvec \in \ \realnum^{\Nfeat}$};
\draw[arrowthick] (cl) -> (z);

\end{tikzpicture}%
}
	\caption{Central image processing: Based on the images, ResNet extracts semantics by classification.}
	\label{fig:resnet_centr}
\end{figure}
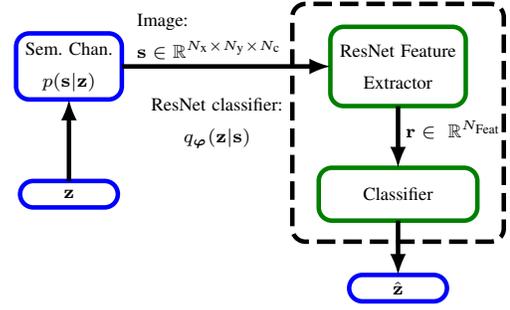

In this section, we provide one numerical example of data-driven semantics to explain what we understand under a semantic communication design and to show its benefits: It is the task of image classification. In fact, we consider our example of the biologist from Sec.~\ref{sec:23} who wants to know which type the tree is.

For the remainder of this article, we will thus assume the hidden semantic \rv\ to be a one-hot vector $\relvec\in \{0,1\}^{\Nclass\times 1}$ where all elements are zero except for one element representing one of the $\Nclass$ image classes. Then, the semantic channel $\prob(\semvec|\relvec)$ (see Fig.~\ref{fig1:com_system_small}) generates images belonging to this class, i.e., the source signal $\semvec$.

Note that for point-to-point transmission as in \cite{shao_learning_2022} we could first classify the image based on the posterior $\aprob_{\rxpars}(\relvec|\semvec)$ as shown in Fig.~\ref{fig:resnet_centr} and transmit the estimate $\relvecest$ (encoded into $\xvec$) through the physical channel since this would be most rate or bandwidth efficient.

But if the image information is distributed across multiple agents, all (sub) images may contribute useful information for classification. We could thus lose information when making hard decisions on each transmitter's side. In the distributed setting, transmission and combination of features, i.e., soft information, is crucial to obtain high classification accuracy.

Further, we note that transmission of full information, i.e., raw image data $\semvec$, through a wireless channel from each agent to a central unit for full image classification would consume a lot of bandwidth. This case is also shown in Fig.~\ref{fig:resnet_centr} assuming perfect communication links between the output of the semantic channel and the input of the ResNet Feature Extractor.

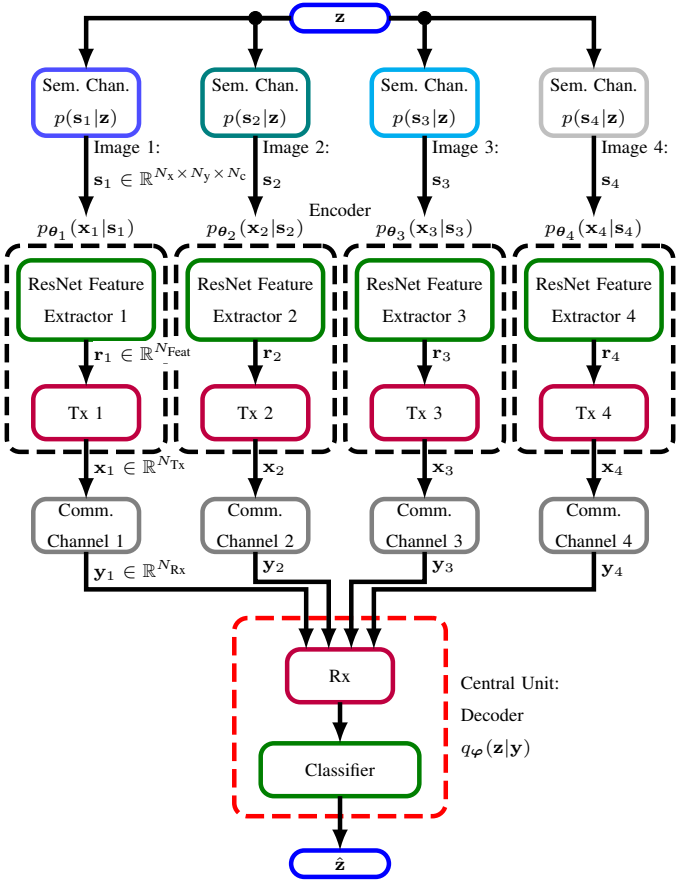
\begin{figure}[!t]
	\centerline{\begin{tikzpicture}[%
					inner xsep=0pt, inner ysep=0pt, %
					outer xsep=0pt, outer ysep=0pt, %
					>=latex,
					ucircle/.style={circle,line width=0.30mm,minimum size=4.00mm,draw=black,fill=white},
					urec/.style={rectangle,line width=0.60mm,draw=black,fill=white,minimum width=50.0,minimum height=20.00},
					invrec/.style={rectangle,line width=0.30mm,draw=none,fill=none,minimum width=50.0,minimum height=20.00},
					arrowthick/.style={->,line width=0.6mm},
					arrowthin/.style={<->,line width=0.6mm},
					linesplit/.style={circle, draw=black, fill=black, inner sep=0pt, minimum height=2},
					font={\fontsize{7pt}{12}\selectfont},
					]

\newcommand{\clsem}{blue} 			%
\newcommand{\clsemchan}{blue!70} 	%
\newcommand{\clenc}{black} 			%
\newcommand{\clresnet}{\clamp} 	%
\newcommand{\clchan}{gray} 			%
\newcommand{\cltrans}{purple} 		%
\newcommand{\cldec}{red} 				%

\node[urec, draw = \clsem, minimum height=10.00, minimum width=10.0, text width=1.3cm, align=center, rounded corners=5] (z0) at (-10.125,1.25) {$\relvec$};
\node[] (Enclab) at (-10.125,-1.3) {Encoder};

\node[linesplit, minimum height=5] (rvnode1) at (-11.25,1.25) {};
\node[linesplit, minimum height=5] (rvnode2) at (-9,1.25) {};

\node[urec, draw = \clsemchan, minimum height = 25.0, minimum width=40.0, text width=1.3cm, align=center, rounded corners=5] (image1) at (-13.5,0.125) {Sem. Chan. $\prob(\semvec_1|\relvec)$};
\node[urec, draw=\clamp, minimum height=30.00, minimum width=50.0, text width=1.8cm, align=center, rounded corners=5] (resnet1) at (-13.5,-2.5) {ResNet Feature Extractor 1};
\node[urec, draw=\cltrans, minimum height=20.00, minimum width=40.0, text width=1.3cm, align=center, rounded corners=5] (tx1) at (-13.5,-4) {Tx 1};
\node[urec, draw = \clenc, minimum height=78.00, minimum width=60.0, text width=1.8cm, fill=none, dashed, dash pattern={on 7pt off 3pt}, rounded corners=5] (Enc1) at (-13.5,-3.15) {};
\node[anchor = south, inner xsep = 0.00cm, inner ysep = 0.08cm, xshift = 0.0cm, text width=2cm, align=center] at (Enc1.north) {$\prob_{\txpars_1}(\xvec_1|\semvec_1)$};
\node[urec, draw = \clchan, minimum height=20.00, minimum width=40.0, text width=1.3cm, align=center, rounded corners=5] (ch1) at (-13.5,-5.5) {Comm. Channel 1};
\draw[arrowthick] (image1) -> (-13.5,-1.4) node[pos=0.35, rotate=0, align=left, anchor=west, xshift = 0.1cm] {Image 1:\\$\semvec_1 \in \realnum^{\Nimx\times \Nimy\times \Nimc}$};
\draw[arrowthick] (tx1) -> (ch1) node[midway, rotate=0, align=center, anchor=west, xshift = 0.1cm, yshift = 0.05cm] {$\xvec_1 \in \realnum^{\ntx}$};

\node[urec, draw = teal, minimum height = 25.0, minimum width=40.0, text width=1.3cm, align=center, rounded corners=5] (image2) at (-11.25,0.125) {Sem. Chan. $\prob(\semvec_2|\relvec)$};
\node[urec, draw=\clamp, minimum height=30.00, minimum width=50.0, text width=1.8cm, align=center, rounded corners=5] (resnet2) at (-11.25,-2.5) {ResNet Feature Extractor 2};
\node[urec, draw=\cltrans, minimum height=20.00, minimum width=40.0, text width=1.3cm, align=center, rounded corners=5] (tx2) at (-11.25,-4) {Tx 2};
\node[urec, draw = \clenc, minimum height=78.00, minimum width=60.0, text width=1.8cm, fill=none, dashed, dash pattern={on 7pt off 3pt}, rounded corners=5] (Enc2) at (-11.25,-3.15) {};
\node[anchor = south, inner xsep = 0.00cm, inner ysep = 0.08cm, xshift = 0.0cm, text width=2cm, align=center] at (Enc2.north) {$\prob_{\txpars_2}(\xvec_2|\semvec_2)$};
\node[urec, draw = \clchan, minimum height=20.00, minimum width=40.0, text width=1.3cm, align=center, rounded corners=5] (ch2) at (-11.25,-5.5) {Comm. Channel 2};
\draw[arrowthick] (image2) -> (-11.25,-1.4) node[pos=0.35, rotate=0, align=left, anchor=west, xshift = 0.1cm] {Image 2:\\$\semvec_2$};
\draw[arrowthick] (resnet2) -> (tx2) node[midway, rotate=0, align=center, anchor=west, xshift = 0.1cm, yshift = 0.1cm] {$\featvec_2$};
\draw[arrowthick] (tx2) -> (ch2) node[midway, rotate=0, align=center, anchor=west, xshift = 0.1cm] {$\xvec_2$};

\node[urec, draw = \clsemchan, draw = cyan, minimum height = 25.0, minimum width=40.0, text width=1.3cm, align=center, rounded corners=5] (image3) at (-9,0.125) {Sem. Chan. $\prob(\semvec_3|\relvec)$};
\node[urec, draw=\clamp, minimum height=30.00, minimum width=50.0, text width=1.8cm, align=center, rounded corners=5] (resnet3) at (-9,-2.5) {ResNet Feature Extractor 3};
\node[urec, draw=\cltrans, minimum height=20.00, minimum width=40.0, text width=1.3cm, align=center, rounded corners=5] (tx3) at (-9,-4) {Tx 3};
\node[urec, draw = \clenc, minimum height=78.00, minimum width=60.0, text width=1.8cm, fill=none, dashed, dash pattern={on 7pt off 3pt}, rounded corners=5] (Enc3) at (-9,-3.15) {};
\node[anchor = south, inner xsep = 0.00cm, inner ysep = 0.08cm, xshift = 0.0cm, text width=2cm, align=center] at (Enc3.north) {$\prob_{\txpars_3}(\xvec_3|\semvec_3)$};
\node[urec, draw = \clchan, minimum height=20.00, minimum width=40.0, text width=1.3cm, align=center, rounded corners=5] (ch3) at (-9,-5.5) {Comm. Channel 3};
\draw[arrowthick] (image3) -> (-9,-1.4) node[pos=0.35, rotate=0, align=left, anchor=west, xshift = 0.1cm] {Image 3:\\$\semvec_3$};
\draw[arrowthick] (resnet3) -> (tx3) node[midway, rotate=0, align=center, anchor=west, xshift = 0.1cm, yshift = 0.1cm] {$\featvec_3$};
\draw[arrowthick] (tx3) -> (ch3)  node[midway, rotate=0, align=center, anchor=west, xshift = 0.1cm] {$\xvec_3$};

\node[urec, draw = \clsemchan, draw = lightgray, minimum height = 25.0, minimum width=40.0, text width=1.3cm, align=center, rounded corners=5] (image4) at (-6.75,0.125) {Sem. Chan. $\prob(\semvec_4|\relvec)$};
\node[urec, draw=\clamp, minimum height=30.00, minimum width=50.0, text width=1.8cm, align=center, rounded corners=5] (resnet4) at (-6.75,-2.5) {ResNet Feature Extractor 4};
\node[urec, draw=\cltrans, minimum height=20.00, minimum width=40.0, text width=1.3cm, align=center, rounded corners=5] (tx4) at (-6.75,-4) {Tx 4};
\node[urec, draw = \clenc, minimum height=78.00, minimum width=60.0, text width=1.8cm, fill=none, dashed, dash pattern={on 7pt off 3pt}, rounded corners=5] (Enc4) at (-6.75,-3.15) {};
\node[anchor = south, inner xsep = 0.00cm, inner ysep = 0.08cm, xshift = 0.0cm, text width=2cm, align=center] at (Enc4.north) {$\prob_{\txpars_4}(\xvec_4|\semvec_4)$};
\node[urec, draw = \clchan, minimum height=20.00, minimum width=40.0, text width=1.3cm, align=center, rounded corners=5] (ch4) at (-6.75,-5.5) {Comm. Channel 4};
\draw[arrowthick] (image4) -> (-6.75,-1.4) node[pos=0.35, rotate=0, align=left, anchor=west, xshift = 0.1cm] {Image 4:\\$\semvec_4$};
\draw[arrowthick] (resnet4) -> (tx4) node[midway, rotate=0, align=center, anchor=west, xshift = 0.1cm, yshift = 0.1cm] {$\featvec_4$};
\draw[arrowthick] (tx4) -> (ch4) node[midway, rotate=0, align=center, anchor=west, xshift = 0.1cm] {$\xvec_4$};

\draw[arrowthick] (resnet1) -> (tx1) node[midway, rotate=0, align=center, anchor=west, xshift = 0.1cm, yshift = 0.14cm, fill=white, minimum height = 8.0] {$\featvec_1 \in \realnum^{\Nfeat}$};

\draw[arrowthick] (z0) -| (image1);
\draw[arrowthick,-] (z0) -> (rvnode1);
\draw[arrowthick,-] (z0) -> (rvnode2);
\draw[arrowthick] (z0) -| (image4);
\draw[arrowthick] (rvnode1) -> (image2);
\draw[arrowthick] (rvnode2) -> (image3);

\node[urec, draw=\cltrans, minimum height=20.00, minimum width=40.0, text width=1.3cm, align=center, rounded corners=5] (rx) at (-10.125,-7.5) {Rx};
\node[urec, draw = \clresnet, minimum height=20.00, minimum width=60.0, text width=1.3cm, align=center, rounded corners=5] (cl) at (-10.125,-8.75) {Classifier};
\node[urec, draw = \clsem, minimum height=10.00, minimum width=10.0, text width=1.3cm, align=center, rounded corners=5] (z) at (-10.125,-10) {$\relvecest$};
\node[urec, draw = \cldec, minimum height=75.00, minimum width=80.0, text width=2.25cm, fill=none, dashed, dash pattern={on 7pt off 3pt}, rounded corners=5, align = left] (Dec) at (-10.125, -8.05) {};
\node[anchor = west, inner sep = 0.00cm, yshift = 0.0cm, xshift = 0.2cm, text width=1.75cm, align=left] at (Dec.east) {Central Unit: \\ Decoder $\aprob_{\rxpars}(\relvec|\yvec)$};

\draw[arrowthick] (ch1.south) -- (-13.5,-6.35) node[midway, rotate=0, align=center, anchor=west, yshift = 0.0cm, xshift = 0.1cm] {$\yvec_1 \in \realnum^{\Nrx}$} -| ([xshift=-0.45cm]rx.north);
\draw[arrowthick] (ch2.south) -- (-11.25,-6.25) node[midway, rotate=0, align=center, anchor=west, yshift = 0.0cm, xshift = 0.1cm] {$\yvec_2$} -| ([xshift=-0.15cm]rx.north);
\draw[arrowthick] (ch3.south) -- (-9.0,-6.25) node[midway, rotate=0, align=center, anchor=west, yshift = 0.0cm, xshift = 0.1cm] {$\yvec_3$} -| ([xshift=0.15cm]rx.north);
\draw[arrowthick] (ch4.south) -- (-6.75,-6.35) node[midway, rotate=0, align=center, anchor=west, yshift = 0.0cm, xshift = 0.1cm] {$\yvec_4$} -| ([xshift=0.45cm]rx.north);

\draw[arrowthick] (rx) -> (cl);
\draw[arrowthick] (cl) -> (z);

\end{tikzpicture}%
}
	\caption{Semantic INFOrmation traNsmission and recoverY (\sinfoni) for distributed agents. Each agent extracts features for bandwidth-efficient transmission. Based on the received signal, the central unit extracts semantics by classification.}
	\label{fig:resnet_distr}
\end{figure}

Therefore, we investigate a distributed setting shown in Fig.~\ref{fig:resnet_distr}. There, each of four agents sees its own image $\semvec_1,\dots,\semvec_4 \sim\prob(\semvec_{\indi}|\relvec)$ being generated by the same semantic \rv\ $\relvec$. Based on these images, a central unit shall extract semantics, i.e., perform classification. We propose to optimize the four encoders $\prob_{\txpars_{\indi}}(\xvec_{\indi}|\semvec_{\indi})$ with $\indi=1,\dots,4$, each consisting of a bandwidth efficient feature extractor (ResNet Feature Extractor $\indi$) and transmitter (Tx $\indi$) \textbf{jointly} with a decoder $\aprob_{\rxpars}(\relvec|\yvec=[\yvec_1,\yvec_2,\yvec_3,\yvec_4]^{\trapo})$, consisting of a receiver (Rx) and concluding classifier (Classifier), w.r.t. cross entropy \eqref{eq:relvar_infomax} of the semantic labels (see Fig.~\ref{fig:resnet_distr}). Hence, we maximize the system's overall semantic measure, i.e., classification accuracy. Note that this scenario is different from both \cite{aguerri_distributed_2021, shao_task-oriented_2023}: We include a physical communication channel (Comm. Channel $\indi$) since we aim to transmit and not only compress. For the sake of simplicity, we assume orthogonal channel access. The IB is addressed by limiting the number of channel uses, which defines the constraint $\mic$ in \eqref{eq:ibm}.

As a first demonstration example, we use the grayscale MNIST and colored CIFAR10 datasets with $\Nclass=10$ image classes \cite{he_deep_2016}. We assume that the semantic channel generates an image that we divide into four equally sized quadrants and each agent observes one quadrant $\semvec_1,\dots,\semvec_4 \in \realnum^{\Nimx\times\Nimy\times\Nimc}$ where $\Nimx$ and $\Nimy$ is the number of image pixels in the x- and y-dimension, respectively, and $\Nimc$ is the number of color channels. Albeit this does not resemble a realistic scenario, note that we can still show the basic working principle and ease implementation.

\subsection{ResNet}

For the design of the overall system, we rely on a famous DNN approach for feature extraction, breaking records at the time of invention: ResNet \cite{he_deep_2016, he_identity_2016}. The key idea of ResNet is that it consists of multiple residual units: Each unit's input is fed directly to its output and if the dimensions do not match, a convolutional layer is used. This structure allows for fast training and convergence of DNNs since the training error can be backpropagated to early layers through these skip connections. From a mathematical point of view, usual DNNs have the design flaw that using a larger function class, i.e., more DNN layers, does not necessarily increase the expressive power. However, this holds for nested functions like ResNet which contain the smaller classes of early layers.

Each residual unit itself consists of two Convolutional NNs (CNNs) with subsequent batch normalization and ReLU activation function, i.e., $\relu=\max(\cdot,0)$, to extract translation invariant and local features across two spatial dimensions $\Nimx$ and $\Nimy$. Color channels like in CIFAR10 add a third dimension $\Nimc=3$ and additional information. The idea behind stacking multiple layers of CNNs is that features tend to become more abstract from early layers (e.g., edges and circles) to final layers (e.g., beaks or tires).

\begin{table}[!t]
	\renewcommand{\arraystretch}{1.3}
	\caption{Semantic INFOrmation TraNsmission and RecoverY (\sinfoni) - DNN architecture for distributed image classification.}
	\label{tab2:com_sem}
	\centering
	\begin{tabular}{lll}
		\hline
		Component	& Layer 					& Dimension								\\
		\hline
		Input		& Image (MNIST, CIFAR10)	& $(14, 14, 1)$, $(16, 16, 3)$			\\
		\hline
		$4\times$	& Conv2D 					& $(14, 14, 14)$, $(16, 16, 16)$		\\
		Feature		& ResNetBlock (2/3 res. un.)	& $(14, 14, 14)$, $(16, 16, 16)$	\\
		Extractor	& ResNetBlock (2/3 res. un.)	& $(7, 7, 28)$, $(8, 8, 32)$		\\
					& ResNetBlock (2/3 res. un.)	& $(4, 4, 56)$, $(4, 4, 64)$		\\
					& Batch Normalization		& $(4, 4, 56)$, $(4, 4, 64)$			\\
					& ReLU activation			& $(4, 4, 56)$, $(4, 4, 64)$			\\
					& GlobalAvgPool2D			& $(56)$, $(64)$						\\
		\hline
		$4\times$ Tx& ReLU						& $\ntx$			\\
					& Linear					& $\ntx$			\\
					& Normalization (dim.)		& $\ntx$			\\
		\hline
		Channel 	& AWGN						& $\ntx$			\\
		\hline
		Rx 			& ReLU ($4\times$ shared)	& $(2, 2, \nrx)$	\\
					& GlobalAvgPool2D			& $\nrx$			\\
		\hline
		Classifier	& Softmax					& $\Nclass=10$		\\
		\hline
	\end{tabular}
\end{table}

In this work, we use the preactivation version of ResNet without bottlenecks from \cite{he_deep_2016, he_identity_2016} implemented for classification on the dataset CIFAR10. In Tab.~\ref{tab2:com_sem}, we show its structure for the distributed scenario from Fig.~\ref{fig:resnet_distr}. There, ResNetBlock is the basic building block of the ResNet architecture. Each block consists of multiple residual units (res. un.) and we use $2$ for the MNIST dataset and $3$ for the CIFAR10 dataset, which means we use ResNet14 and ResNet20, respectively. We arrive at the architecture of central image processing from Fig.~\ref{fig:resnet_centr} by removing the components Tx, (physical) Channel, and Rx and increasing each spatial dimension by $2$ to contain all quadrants of the original image. For further implementation details, we refer the reader to the original work \cite{he_identity_2016}.

\subsection{Distributed Semantic Communication Design Approach}

Our key idea here is to modify ResNet w.r.t. the communication task by splitting it at a suitable point where a representation $\featvec\in\realnum^{\Nfeat\times 1}$ of semantic information with low-bandwidth is present (see Fig.~\ref{fig:resnet_centr} and \ref{fig:resnet_distr}). ResNet and CNNs in general can be interpreted to extract features: With full images, we obtain a feature map of size $8\times 8\times \Nfeat$ after the last ReLU activation (see Tab.~\ref{tab2:com_sem}). These local features are aggregated by the global average pooling layers across the $2$ spatial dimensions into $\featvec$. Based on these $\Nfeat$ global features in $\featvec$, the softmax layer finally classifies the image. We note that the features contain the relevant information w.r.t. the semantic \rv\ $\relvec$ and are of low dimension compared to the original image or even its sub-images, i.e., $64$ compared to $16\times 16\times 3=768$ for CIFAR10.

Therefore, we aim to transmit each agent's local features $\featvec_{\indi}\in\realnum^{\Nfeat\times 1}$ ($\indi=1,\dots,4$) instead of all sub-images $\semvec_{\indi}$ and add the component Tx in Tab.~\ref{tab2:com_sem} to encode the features $\featvec_{\indi}$ into $\xvec_{\indi}\in\realnum^{\ntx\times 1}$ for transmission through the wireless channel (see Fig.~\ref{fig:resnet_distr}). We note that $\xvec_{\indi}\in\realnum^{\ntx\times 1}$ is analog and that the output dimension $\ntx$ of $\xvec_{\indi}$ defines the number of channel uses per agent/image. Note that the less often we use the wireless channel ($\ntx$), the less information we transmit but the less bandwidth we consume, and vice versa. Hence, the number of channel uses defines the IB in \eqref{eq:ibm}. We implement the Tx module by DNN layers. To limit the transmit power to one, we constrain the Tx output by the norm along the training batch or the encode vector dimension (dim.), e.g., $\xvec_{\indi}=\sqrt{\ntx}\cdot\xvecwnorm_{\indi}/\norm{\xvecwnorm_{\indi}}_2$ where $\xvecwnorm_{\indi}\in\realnum^{\ntx\times 1}$ is the output of the layer Linear from Tab.~\ref{tab2:com_sem}. For numerical simulations, we choose all Tx layers to have width $\ntx$.

At the receiver side, we use a single Rx module only with shared DNN layers and parameters $\rxpars_{\txt{Rx}}$ for all inputs $\yvec_{\indi}$: This setting would be optimal if any feature is reflected in any sub-image and if the statistics of the physical channels are the same. Exploiting the prior knowledge of location-invariant features and assuming Additive White Gaussian Noise (AWGN) channels, this design choice seems reasonable. In our experiments, all layers of the Rx module have width $\nrx$. A larger layer width $\nrx$ is equivalent to more computing power.

The output of the Rx module can be interpreted as a representation of the image features $\featvec_{\indi}$ with index $\indi$ indicating the spatial location. Thus, we have a representation of a feature map of size $(2, 2, \nrx)$ that we aggregate across the spatial dimension according to the ResNet structure. Based on this semantic representation, a softmax layer with $10$ units finally computes class probabilities $\aprob_{\rxpars}(\relvec|\yvec)$ whose maximum is the maximum a posteriori estimate $\relvarest$. In the following, we name our proposed approach Semantic INFOrmation TraNsmission and RecoverY (\sinfoni).

\subsection{Optimization Details}
\label{sec:33}

We evaluate \sinfoni\ in TensorFlow 2 \cite{martin_abadi_tensorflow_2015} on the MNIST and CIFAR10 datasets. The source code is available in \cite{Beck_Semantic_INFOrmation_traNsmission_2023}. We split the dataset into $60$k/$50$k training data and $10$k validation data samples, respectively. For preprocessing, we normalize the pixel inputs to range $[0,1]$, but we do not use data augmentation, in contrast to \cite{he_deep_2016, he_identity_2016}, yielding slightly worse accuracy. The ReLU layers are initialized with uniform distribution according to He and all other layers according to Glorot \cite{he_delving_2015}.

In the case of CIFAR10 classification with central image processing and original ResNet, we need to train $\Ntxpars+\Nrxpars=273,066$ parameters. We like to stress that although we divided the image input into four smaller pieces, this number grows more than four times to $4\Ntxpars+\Nrxpars=1,127,754$ with $\ntx=\Nfeat=64$ for \sinfoni. The reason lies in the ResNet structure with minor dependence on the input image size and that we process at four agents with an additional Tx module. Only $\Nrxpars=4,810$ parameters amount to the Rx module and classification, i.e., the central unit. We note that the number of added Tx and Rx parameters of $33,560$ and $3,192$ is relatively small. Since the number of parameters only weakly grows with Rx layer width $\nrx$ in our design, we choose $\nrx=\Nfeat$ as the default.

For optimization of the cross entropy \eqref{eq:relvar_infomax} or the loss function \eqref{eq:mcloss}, we use the reparametrization trick from Sec.~\ref{sec:3impl} and Stochastic Gradient Descent (SGD) with a momentum of $0.9$ and a batch size of $\Nb=64$. We add $\lnorm[2]$-regularization with a weight decay of $0.0001$ as in \cite{he_deep_2016, he_identity_2016}. The learning rate of $\lr=0.1$ is reduced to $0.01$ and $0.001$ after $\Ne=100$ and $150$ epochs for CIFAR10 and after $3$ and $6$ for MNIST. In total, we train for $\Ne=200$ epochs with CIFAR10 and for $20$ with MNIST. In order to optimize the transceiver for a wider SNR range, we choose the SNR to be uniformly distributed within $[-4,6]$ dB where $\txt{SNR}=1/\noisestd^2$ with noise variance $\noisestd^2$.

\subsection{Numerical Results}

In the following, we will investigate the influence of specific design choices on our semantic approach \sinfoni. Then, we compare a semantic with a classical Shannon-based transmission approach. The design choices are as follows:
\begin{itemize}
	\item \textbf{Central:} Central and joint processing of full image information by ResNet classifier, see Fig.~\ref{fig:resnet_centr}. It indicates the maximum achievable accuracy.
	\item \textbf{\sinfoniperfcomm:} The proposed distributed design \sinfoni\ trained with perfect communication links and without channel encoding, i.e., Tx and Rx module, but with Tx normalization layer. Thus, the plain and power-constrained features are transmitted with $\ntx=\Nfeat$ channel uses. It serves as the benchmark since it indicates the maximum performance of the distributed design.
	\item \textbf{\sinfonichan:} \sinfoni\ perfect comm. evaluated with AWGN channel.
	\item \textbf{\sinfonichantrain:} \sinfoni\ perfect comm. trained with AWGN channel.
	\item \textbf{\sinfonitxrx\ ($\bog{\ntx=\Nfeat}$):} \sinfoni\ trained with channel encoding, i.e., Tx and Rx module, and $\ntx=\Nfeat$ channel uses.
	\item \textbf{\sinfonitxrx\ ($\bog{\ntx<\Nfeat}$):} \sinfoni\ trained with channel encoding and $\ntx<\Nfeat$ channel uses for feature compression.
	\item \textbf{\sinfoniclassdig:} \sinfoniperfcomm\ with classic digital communications (Huffman coding, LDPC coding with belief propagation decoding, and BPSK modulation) as additional Tx and Rx modules. For details, see Sec.~\ref{sec:semvsclassic}.
	\item \textbf{\sinfoniclassanae:} \sinfoniperfcomm\ with ML-based analog communications (AE for any element in $\featvec_{\indi}$) as additional Tx and Rx modules. It is basically the semantic communication approach from \cite{farsad_deep_2018, bourtsoulatze_deep_2019, beck_swarm_2023}. For details, see Sec.~\ref{sec:semvsae}.
\end{itemize}
Since meaning is expressed by the \rv\ $\relvec$, we use classification accuracy to measure semantic transmission quality. For illustration in logarithmic scale, we show the opposite of accuracy in all plots, i.e., classification error rate.

\subsubsection{MNIST dataset}

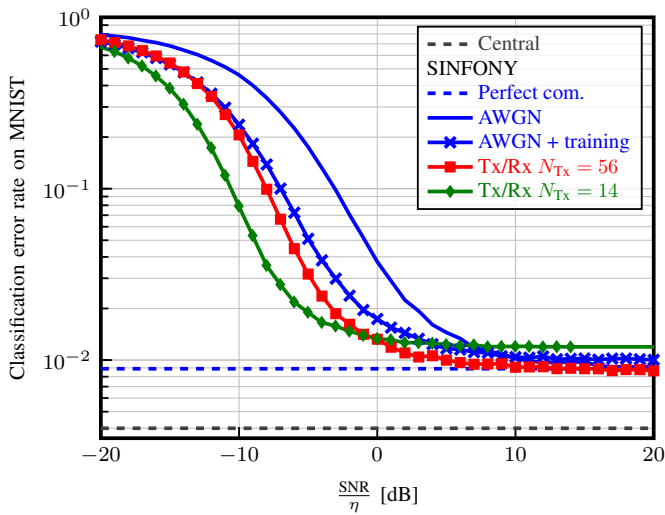
\begin{figure}[!t]
	\centerline{\begin{tikzpicture}%
\begin{axis}[
width=10.5cm,
scale=0.7,
scale only axis,
xlabel={$\frac{\snr}{\speceff}$ [dB]},
xmajorgrids,
xmin=-20, xmax=20,
xminorgrids,
ylabel={Classification error rate on MNIST},
ymajorgrids,
ymin=3.5e-3, ymax=1,
yminorgrids,
ymode=log,
axis background/.style={fill=white},
legend entries={{\color{\clsd} \centralclass},
				{\hspace{-1.075cm} \sinfoni},
				{\color{\cloamp}\perfcomm},
				{\color{\cloamp}AWGN},
				{\color{\cloamp}AWGN + training},
				{\color{\clcmd}\txrx\ $\ntx=56$},
				{\color{\clamp}\txrx\ $\ntx=14$}, %
				},
]
\addplot [ptsd, dashed, mark = None, mark options={solid}]
table {%
-20 0.00400000810623169
20 0.00400000810623169
};
\addplot [ptoamp, empty legend, dashed, mark = None, mark options={solid}]
table {%
0 0
1 2
};
\addplot [ptoamp, dashed, mark = None, mark options={solid}]
table {%
-20 0.008899986743927
20 0.008899986743927
};
\addplot [ptoamp, mark = None, mark options={solid}]
table {%
-30 0.872280000150204
-29 0.867200000584125
-28 0.861020001769066
-27 0.856790001690388
-26 0.851319999992847
-25 0.845340000092983
-24 0.838970001041889
-23 0.830439999699593
-22 0.818590001761913
-21 0.805629998445511
-20 0.792959998548031
-19 0.77609000056982
-18 0.760259999334812
-17 0.734090000391006
-16 0.71186999976635
-15 0.68129999935627
-14 0.642899996042252
-13 0.605340003967285
-12 0.559689998626709
-11 0.512590003013611
-10 0.460640001296997
-9 0.400339996814728
-8 0.338619995117188
-7 0.279580003023148
-6 0.224939996004105
-5 0.175160002708435
-4 0.131659990549088
-3 0.098089998960495
-2 0.0703199982643128
-1 0.0516199946403504
0 0.0374700009822846
1 0.0289499998092652
2 0.0223900020122528
3 0.0193099915981293
4 0.0160600006580353
5 0.0144899964332581
6 0.0133600056171417
7 0.0117799997329712
8 0.0112800002098083
9 0.0108700096607208
10 0.0102200031280517
11 0.00988999605178831
12 0.00966999530792234
13 0.00972999334335323
14 0.00974999666213994
15 0.00971999764442444
16 0.00927001237869263
17 0.00915000438690183
18 0.00914000272750859
19 0.00910000205039974
20 0.00925000309944157
};
\addplot [ptoamp]
table {%
-30 0.856520000100136
-29 0.84980999827385
-28 0.840700000524521
-27 0.833349999785423
-26 0.824689999222755
-25 0.812890000641346
-24 0.800229999423027
-23 0.785349999368191
-22 0.767540000379086
-21 0.747940003871918
-20 0.722040003538132
-19 0.694560000300407
-18 0.659970000386238
-17 0.625310003757477
-16 0.579249992966652
-15 0.53105999827385
-14 0.477259999513626
-13 0.416799998283386
-12 0.358309996128082
-11 0.297049993276596
-10 0.236100000143051
-9 0.182960003614426
-8 0.138460004329681
-7 0.100139993429184
-6 0.0724400043487549
-5 0.0509700059890748
-4 0.0381699979305268
-3 0.0298699915409087
-2 0.0237600028514862
-1 0.0195999920368195
0 0.0173700034618379
1 0.0154700040817259
2 0.0144100069999695
3 0.0133000016212463
4 0.0123800039291381
5 0.0116800010204314
6 0.0115099966526031
7 0.0111299991607665
8 0.0110300004482269
9 0.0105299949645997
10 0.0104099929332732
11 0.0103299915790557
12 0.0104999959468842
13 0.0101099908351899
14 0.0101700007915497
15 0.0102100133895874
16 0.010050004720688
17 0.0101999998092651
18 0.010199999809265
19 0.0101099967956542
20 0.0100300014019012
};
\addplot [ptcmd]
table {%
-30 0.862450003623962
-29 0.858670000731945
-28 0.853880000114441
-27 0.84428000152111
-26 0.834960001707077
-25 0.826350000500679
-24 0.815559999644756
-23 0.799390000104904
-22 0.784559999406338
-21 0.763169999420643
-20 0.739959993958473
-19 0.712070000171661
-18 0.680809995532036
-17 0.639479997754097
-16 0.594949996471405
-15 0.541249999403954
-14 0.479689997434616
-13 0.410810005664825
-12 0.345440012216568
-11 0.270700007677078
-10 0.205979990959168
-9 0.144449996948242
-8 0.099260002374649
-7 0.0663699924945831
-6 0.0446700036525727
-5 0.0317500054836273
-4 0.0236199855804443
-3 0.0186600029468537
-2 0.0161800026893616
-1 0.014139997959137
0 0.0132299959659576
1 0.0118799984455109
2 0.0110099911689758
3 0.0104199945926666
4 0.0105999946594239
5 0.0099900007247925
6 0.0096700012683868
7 0.00945001244544985
8 0.00949000120162968
9 0.00952000021934507
10 0.00908000469207759
11 0.00912000536918645
12 0.00914000272750859
13 0.00891000628471372
14 0.00896001458168028
15 0.00890001058578493
16 0.00888999700546267
17 0.00867000818252561
18 0.00879999995231628
19 0.00879001021385195
20 0.00865001082420347
};
\addplot [ptamp]
table {%
-36.0205999132796 0.883469999581575
-35.0205999132796 0.878139999508858
-34.0205999132796 0.874419999867678
-33.0205999132796 0.873889999091625
-32.0205999132796 0.866960000991821
-31.0205999132796 0.862809999287128
-30.0205999132796 0.854310001432896
-29.0205999132796 0.847750002145767
-28.0205999132796 0.838139997422695
-27.0205999132796 0.829299999773502
-26.0205999132796 0.817570000886917
-25.0205999132796 0.802169999480247
-24.0205999132796 0.785279996693134
-23.0205999132796 0.765709999203682
-22.0205999132796 0.739609995484352
-21.0205999132796 0.710900002717972
-20.0205999132796 0.670579996705055
-19.0205999132796 0.632500001788139
-18.0205999132796 0.579270005226135
-17.0205999132796 0.519959998130798
-16.0205999132796 0.454869997501373
-15.0205999132796 0.385969990491867
-14.0205999132796 0.311049997806549
-13.0205999132796 0.237619996070862
-12.0205999132796 0.17338000535965
-11.0205999132796 0.119789999723434
-10.0205999132796 0.0793000042438509
-9.02059991327963 0.0532999932765961
-8.02059991327963 0.0357299983501433
-7.02059991327962 0.0275699973106384
-6.02059991327962 0.0218200027942658
-5.02059991327962 0.0190099954605102
-4.02059991327962 0.0165800094604491
-3.02059991327962 0.0158100008964539
-2.02059991327962 0.0147600054740906
-1.02059991327962 0.0139400005340575
-0.0205999132796242 0.0133400022983551
0.979400086720376 0.0131099939346314
1.97940008672038 0.0126400053501129
2.97940008672038 0.0126300036907196
3.97940008672038 0.0125299990177153
4.97940008672038 0.0123200058937073
5.97940008672038 0.012099987268448
6.97940008672038 0.0122699916362763
7.97940008672038 0.0119500041007996
8.97940008672037 0.0119999945163727
9.97940008672037 0.0120099961757659
10.9794000867204 0.0120299994945525
11.9794000867204 0.011930000782013
12.9794000867204 0.0119700074195861
13.9794000867204 0.0119400084018709
21 0.0119400084018709
};
\end{axis}

\end{tikzpicture}}
	\caption{Classification error rate of different \sinfoni\ examples (distributed setting) and central image processing on the MNIST validation dataset as a function of normalized SNR.}
	\label{fig:sem_mnist}
\end{figure}

The numerical results of our proposed approach \sinfoni\ on the MNIST validation dataset are shown in Fig.~\ref{fig:sem_mnist} for $\nrx=56$. To obtain a fair comparison between transmit signals $\xvec_{\indi}\in\realnum^{\ntx\times 1}$ of different length $\ntx$, we normalize the SNR by the spectral efficiency or rate $\speceff=\Nfeat/\ntx$. First, we observe that the classification error rate of $0.5\%$ of the central ResNet unit with full image information (Central) is smaller than that of $0.9\%$ of \sinfoniperfcomm. Note that we assume ideal communication links. However, the difference seems negligible considering that the local agents only see a quarter of the full images and learn features independently based on it.

With noisy communication links (\sinfonichan), the performance degrades especially for $\txt{SNR}<10$ dB, and we can avoid degradation just partly by training with noise (\sinfonichantrain). Introducing the Tx module (\sinfonitxrx\ $\ntx=56$), we further improve classification accuracy at low SNR. %
If we encode the features from $\Nfeat=56$ to only $\ntx=14$ in the Tx module (\sinfonitxrx\ $\ntx=14$) to have less channel uses/bandwidth (stronger bottleneck), the error rate is lowest compared to other \sinfoni\ examples with non-ideal links for low normalized SNR. At high SNR, we observe a small error offset, which indicates lossy compression. In fact, our system \sinfoni\ learns a reliable semantic encoding to improve the classification performance of the overall system with non-ideal links. Every design choice in Tab.~\ref{tab2:com_sem} is well-motivated.

\subsubsection{CIFAR10 dataset}

\begin{figure}[!t]
	\centerline{\begin{tikzpicture}%
\begin{axis}[
width=10.5cm,
scale=0.7,
scale only axis,
xlabel={$\frac{\snr}{\speceff}$ [dB]},
xmajorgrids,
xmin=-20, xmax=20,
xminorgrids,
ylabel={Classification error rate on CIFAR10},
ymajorgrids,
ymin=1e-1, ymax=1,
yminorgrids,
ymode=log,
axis background/.style={fill=white},
legend entries={{\color{\clsd} \centralclass},
				{\hspace{-1.075cm} \sinfoni},
				{\color{\cloamp}\perfcomm},
				{\color{\cloamp}AWGN},
				{\color{\cloamp}AWGN + training}, %
				{\color{\clcmd}\txrx\ $\ntx=64$}, %
				{\color{\clamp}\txrx\ $\ntx=16$}, %
				},
]
\addplot [ptsd, dashed, mark = None, mark options={solid}]
table {%
-20 0.116699993610382
20 0.116699993610382
};
\addplot [ptoamp, empty legend, dashed, mark = None, mark options={solid}]
table {%
0 0
1 2
};
\addplot [ptoamp, dashed, mark = None, mark options={solid}]
table {%
-20 0.198400020599365
20 0.198400020599365
};
\addplot [ptoamp, mark = None, mark options={solid}]
table {%
-30 0.884389999508858
-29 0.881300000846386
-28 0.878150000423193
-27 0.877460002154112
-26 0.873710000514984
-25 0.868659999966621
-24 0.864170001447201
-23 0.858769999444485
-22 0.854210002720356
-21 0.847779999673367
-20 0.839569999277592
-19 0.832209998369217
-18 0.822359998524189
-17 0.812070000171661
-16 0.797990000247955
-15 0.784640000760555
-14 0.765139999985695
-13 0.746240000426769
-12 0.72248999774456
-11 0.699559995532036
-10 0.666549998521805
-9 0.635499995946884
-8 0.599969998002052
-7 0.562909996509552
-6 0.527029997110367
-5 0.485279995203018
-4 0.447089993953705
-3 0.410680001974106
-2 0.378689992427826
-1 0.349009996652603
0 0.326480001211166
1 0.301709997653961
2 0.2832200050354
3 0.265330004692078
4 0.254559993743896
5 0.242979991436005
6 0.234699994325638
7 0.225889992713928
8 0.221839994192123
9 0.217370003461838
10 0.212580001354217
11 0.210659998655319
12 0.207639992237091
13 0.206279999017715
14 0.204449993371963
15 0.205029994249344
16 0.203199994564056
17 0.202660006284714
18 0.201579999923706
19 0.200580006837845
20 0.200900000333786
};
\addplot [ptoamp]
table {%
-30 0.870920000225305
-29 0.866140000522137
-28 0.861370000243187
-27 0.856189997494221
-26 0.84890999943018
-25 0.843350000679493
-24 0.835959999263287
-23 0.827679999172688
-22 0.814390000700951
-21 0.801899999380112
-20 0.784910000860691
-19 0.770199999213219
-18 0.750310002267361
-17 0.723129999637604
-16 0.695980000495911
-15 0.665650004148483
-14 0.635290002822876
-13 0.599289998412132
-12 0.55779000222683
-11 0.522039994597435
-10 0.48235000371933
-9 0.440670001506805
-8 0.407100003957748
-7 0.372860008478165
-6 0.347030001878738
-5 0.325899994373322
-4 0.30819000005722
-3 0.292249995470047
-2 0.281110000610352
-1 0.271079993247986
0 0.264419996738434
1 0.259410005807877
2 0.255049997568131
3 0.251700007915497
4 0.248309999704361
5 0.245390003919601
6 0.24465999007225
7 0.242190003395081
8 0.241549998521805
9 0.239700001478195
10 0.238610005378723
11 0.238309997320175
12 0.237910002470016
13 0.238109993934631
14 0.236999994516373
15 0.236669999361038
16 0.237030005455017
17 0.23604000210762
18 0.236000001430511
19 0.236299997568131
20 0.235969996452332
};
\addplot [ptcmd]
table {%
-30 0.872039999812841
-29 0.866889999806881
-28 0.862209998071194
-27 0.856180000305176
-26 0.849659998714924
-25 0.841160000860691
-24 0.833710001409054
-23 0.823780001699924
-22 0.809149999916553
-21 0.795589998364449
-20 0.776590000092983
-19 0.756219999492168
-18 0.72960000038147
-17 0.699909999966621
-16 0.664319998025894
-15 0.622240000963211
-14 0.577520000934601
-13 0.532890000939369
-12 0.481240004301071
-11 0.42849001288414
-10 0.38357999920845
-9 0.342030000686646
-8 0.306390011310577
-7 0.278549993038178
-6 0.25621999502182
-5 0.241369986534119
-4 0.230580008029938
-3 0.221750003099441
-2 0.216449993848801
-1 0.210920000076294
0 0.206949996948242
1 0.205580002069473
2 0.203139996528626
3 0.202050000429153
4 0.200330001115799
5 0.200669997930527
6 0.199090003967285
7 0.198720002174377
8 0.198710006475449
9 0.198779994249344
10 0.198199999332428
11 0.19860999584198
12 0.198229998350143
13 0.198330003023148
14 0.197909992933273
15 0.197710001468658
16 0.197420001029968
17 0.197689998149872
18 0.197900009155274
19 0.197340005636215
20 0.197319996356964
};
\addplot [ptamp]
table {%
-35.4406804435028 0.87470999956131
-34.4406804435028 0.872859997302294
-33.4406804435028 0.86697999984026
-32.4406804435028 0.864339996874332
-31.4406804435028 0.858669997751713
-30.4406804435028 0.853350001573563
-29.4406804435028 0.844919998943806
-28.4406804435028 0.836540001630783
-27.4406804435028 0.828499999642372
-26.4406804435028 0.813620001077652
-25.4406804435028 0.800770001113415
-24.4406804435028 0.783859999477863
-23.4406804435028 0.76826000213623
-22.4406804435028 0.742249992489815
-21.4406804435028 0.717150005698204
-20.4406804435028 0.687270000576973
-19.4406804435028 0.652360007166862
-18.4406804435028 0.613390001654625
-17.4406804435028 0.57296000123024
-16.4406804435028 0.528100001811981
-15.4406804435028 0.481629991531372
-14.4406804435028 0.434729999303818
-13.4406804435028 0.393640005588532
-12.4406804435028 0.355889999866486
-11.4406804435028 0.319499999284744
-10.4406804435028 0.293590003252029
-9.44068044350276 0.272330003976822
-8.44068044350276 0.254800009727478
-7.44068044350276 0.243370008468628
-6.44068044350276 0.231270003318787
-5.44068044350276 0.225409996509552
-4.44068044350276 0.220260000228882
-3.44068044350276 0.215230000019073
-2.44068044350276 0.213670003414154
-1.44068044350276 0.210820001363754
-0.440680443502757 0.208679991960525
0.559319556497243 0.207970005273819
1.55931955649724 0.20595999956131
2.55931955649724 0.205930000543594
3.55931955649724 0.205110001564026
4.55931955649724 0.204350000619888
5.55931955649724 0.203970003128052
6.55931955649724 0.203249996900559
7.55931955649724 0.203769999742508
8.55931955649724 0.203010004758835
9.55931955649724 0.20310999751091
10.5593195564972 0.203320002555847
11.5593195564972 0.202840000391006
12.5593195564972 0.202820003032684
13.5593195564972 0.202849990129471
14.5593195564972 0.202610003948212
21 0.202610003948212
};
\end{axis}

\end{tikzpicture}}
	\caption{Classification error rate of different \sinfoni\ examples (distributed setting) and central image processing on CIFAR10 as a function of normalized SNR.}
	\label{fig:sem_cifar}
\end{figure}
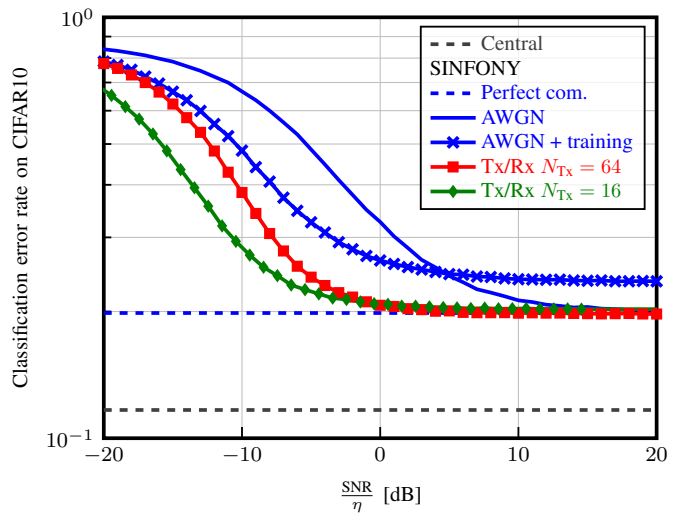

Comparing these results to the classification accuracy on CIFAR10 shown in Fig.~\ref{fig:sem_cifar}, we observe a similar behavior. But a few main differences become apparent: Central performs much better with $12\%$ error rate than \sinfoniperfcomm\ with $20\%$. We expect the reason to lie in the more challenging dataset with more color channels. Further, \sinfonichantrain\ with $\ntx=\Nfeat=64$ channel uses runs into a rather high error floor. Notably, even \sinfonitxrx\ ($\ntx=16$) with fewer channel uses performs better than both \sinfonichan\ and \sinfonichantrain\ over the whole SNR range and achieves channel encoding with negligible loss. This means adding more flexible channel encoding, i.e., Tx/Rx module, is crucial for CIFAR10.

\subsubsection{Channel Uses Constraint}

\begin{figure}[!t]
	\centerline{\begin{tikzpicture}%
\begin{axis}[
width=10.5cm,
scale=0.7,
scale only axis,
xlabel={$\frac{\snr}{\speceff}$ [dB]},
xmajorgrids,
xmin=-18, xmax=20,
xminorgrids,
ylabel={Classification error rate on MNIST},
ymajorgrids,
ymin=7e-3, ymax=1,
yminorgrids,
ymode=log,
axis background/.style={fill=white},
legend entries={%
				{\hspace{-1.075cm} \sinfoni},
				{\color{\cloamp}\perfcomm},				
				{\color{\clcmd}\txrx\ $N_{\text{Tx}}=56$},
				{\color{\clamp}\txrx\ $N_{\text{Tx}}=14$},
				{\color{\clamp}\txrx\ $N_{\text{Tx}}=7$},
				{\color{\clamp}\txrx\ $N_{\text{Tx}}=4$},
				{\color{\clamp}\txrx\ $N_{\text{Tx}}=2$},
				},
]
\addplot [ptoamp, empty legend, dashed, mark = None, mark options={solid}]
table {%
0 0
1 2
};
\addplot [ptoamp, dashed, mark = None, mark options={solid}]
table {%
-20 0.008899986743927
20 0.008899986743927
};
\addplot [ptcmd]
table {%
-30 0.862450003623962
-29 0.858670000731945
-28 0.853880000114441
-27 0.84428000152111
-26 0.834960001707077
-25 0.826350000500679
-24 0.815559999644756
-23 0.799390000104904
-22 0.784559999406338
-21 0.763169999420643
-20 0.739959993958473
-19 0.712070000171661
-18 0.680809995532036
-17 0.639479997754097
-16 0.594949996471405
-15 0.541249999403954
-14 0.479689997434616
-13 0.410810005664825
-12 0.345440012216568
-11 0.270700007677078
-10 0.205979990959168
-9 0.144449996948242
-8 0.099260002374649
-7 0.0663699924945831
-6 0.0446700036525727
-5 0.0317500054836273
-4 0.0236199855804443
-3 0.0186600029468537
-2 0.0161800026893616
-1 0.014139997959137
0 0.0132299959659576
1 0.0118799984455109
2 0.0110099911689758
3 0.0104199945926666
4 0.0105999946594239
5 0.0099900007247925
6 0.0096700012683868
7 0.00945001244544985
8 0.00949000120162968
9 0.00952000021934507
10 0.00908000469207759
11 0.00912000536918645
12 0.00914000272750859
13 0.00891000628471372
14 0.00896001458168028
15 0.00890001058578493
16 0.00888999700546267
17 0.00867000818252561
18 0.00879999995231628
19 0.00879001021385195
20 0.00865001082420347
};
\addplot [ptamp]
table {%
-36.0205999132796 0.883469999581575
-35.0205999132796 0.878139999508858
-34.0205999132796 0.874419999867678
-33.0205999132796 0.873889999091625
-32.0205999132796 0.866960000991821
-31.0205999132796 0.862809999287128
-30.0205999132796 0.854310001432896
-29.0205999132796 0.847750002145767
-28.0205999132796 0.838139997422695
-27.0205999132796 0.829299999773502
-26.0205999132796 0.817570000886917
-25.0205999132796 0.802169999480247
-24.0205999132796 0.785279996693134
-23.0205999132796 0.765709999203682
-22.0205999132796 0.739609995484352
-21.0205999132796 0.710900002717972
-20.0205999132796 0.670579996705055
-19.0205999132796 0.632500001788139
-18.0205999132796 0.579270005226135
-17.0205999132796 0.519959998130798
-16.0205999132796 0.454869997501373
-15.0205999132796 0.385969990491867
-14.0205999132796 0.311049997806549
-13.0205999132796 0.237619996070862
-12.0205999132796 0.17338000535965
-11.0205999132796 0.119789999723434
-10.0205999132796 0.0793000042438509
-9.02059991327963 0.0532999932765961
-8.02059991327963 0.0357299983501433
-7.02059991327962 0.0275699973106384
-6.02059991327962 0.0218200027942658
-5.02059991327962 0.0190099954605102
-4.02059991327962 0.0165800094604491
-3.02059991327962 0.0158100008964539
-2.02059991327962 0.0147600054740906
-1.02059991327962 0.0139400005340575
-0.0205999132796242 0.0133400022983551
0.979400086720376 0.0131099939346314
1.97940008672038 0.0126400053501129
2.97940008672038 0.0126300036907196
3.97940008672038 0.0125299990177153
4.97940008672038 0.0123200058937073
5.97940008672038 0.012099987268448
6.97940008672038 0.0122699916362763
7.97940008672038 0.0119500041007996
8.97940008672037 0.0119999945163727
9.97940008672037 0.0120099961757659
10.9794000867204 0.0120299994945525
11.9794000867204 0.011930000782013
12.9794000867204 0.0119700074195861
13.9794000867204 0.0119400084018709
21 0.0119400084018709
};
\addplot [ptamp, dashed, mark=x, mark size=3, mark options={solid}] %
table {%
-39.0308998699194 0.883680000156164
-38.0308998699194 0.88163999915123
-37.0308998699194 0.880169999599457
-36.0308998699194 0.876829998940229
-35.0308998699194 0.874790000170469
-34.0308998699194 0.87090999931097
-33.0308998699194 0.867219999432564
-32.0308998699194 0.861309999227524
-31.0308998699194 0.853069999814034
-30.0308998699194 0.847110000252724
-29.0308998699194 0.837479999661446
-28.0308998699194 0.828210002183914
-27.0308998699194 0.81807000041008
-26.0308998699194 0.803869999945164
-25.0308998699194 0.789129999279976
-24.0308998699194 0.770330001413822
-23.0308998699194 0.744270001351833
-22.0308998699194 0.717429998517036
-21.0308998699194 0.687250000238419
-20.0308998699194 0.647849997878075
-19.0308998699194 0.602130001783371
-18.0308998699194 0.547050002217293
-17.0308998699194 0.487540006637573
-16.0308998699194 0.42153999209404
-15.0308998699194 0.351679992675781
-14.0308998699194 0.278850001096725
-13.0308998699194 0.213800001144409
-12.0308998699194 0.155289995670319
-11.0308998699194 0.107499980926514
-10.0308998699194 0.074779999256134
-9.03089986991944 0.0519700109958648
-8.03089986991944 0.0390799999237061
-7.03089986991944 0.0312200009822845
-6.03089986991944 0.0270500004291535
-5.03089986991944 0.0237900018692018
-4.03089986991944 0.0216699957847596
-3.03089986991944 0.0205299973487854
-2.03089986991944 0.0190699994564056
-1.03089986991944 0.0185799956321717
-0.0308998699194358 0.0180400013923646
0.969100130080564 0.0176599979400636
1.96910013008056 0.0173000037670135
2.96910013008056 0.0170300006866455
3.96910013008056 0.0168800055980681
4.96910013008056 0.016630005836487
5.96910013008056 0.0164799988269807
6.96910013008056 0.0166000008583069
7.96910013008056 0.0165700078010561
8.96910013008056 0.0163900077342988
9.96910013008056 0.0164499998092651
10.9691001300806 0.0162200093269349
21 0.0162200093269349
};
\addplot [ptamp, dashed, mark=*, mark options={solid}]
table {%
-41.4612803567824 0.888740000873804
-40.4612803567824 0.88952000066638
-39.4612803567824 0.887770000100136
-38.4612803567824 0.883860000967979
-37.4612803567824 0.883720000088215
-36.4612803567824 0.879969999939203
-35.4612803567824 0.876139999181032
-34.4612803567824 0.873759999871254
-33.4612803567824 0.867550000548363
-32.4612803567824 0.864289999008179
-31.4612803567824 0.861169998347759
-30.4612803567824 0.853230001032352
-29.4612803567824 0.847029998898506
-28.4612803567824 0.836709998548031
-27.4612803567824 0.824300000071526
-26.4612803567824 0.817569997906685
-25.4612803567824 0.800919997692108
-24.4612803567824 0.782919999957085
-23.4612803567824 0.76438000202179
-22.4612803567824 0.738070005178452
-21.4612803567824 0.711309996247292
-20.4612803567824 0.674659997224808
-19.4612803567824 0.63139999806881
-18.4612803567824 0.58308000266552
-17.4612803567824 0.527410000562668
-16.4612803567824 0.466869986057281
-15.4612803567824 0.394490003585815
-14.4612803567824 0.327110004425049
-13.4612803567824 0.260380005836487
-12.4612803567824 0.198060005903244
-11.4612803567824 0.146169996261597
-10.4612803567824 0.105759996175766
-9.46128035678238 0.0773400008678437
-8.46128035678238 0.0593899965286255
-7.46128035678238 0.0473300039768219
-6.46128035678238 0.0404899954795838
-5.46128035678238 0.0367999851703644
-4.46128035678238 0.034170001745224
-3.46128035678238 0.0312400162220001
-2.46128035678238 0.0308100044727325
-1.46128035678238 0.0302500128746033
-0.461280356782382 0.0292200088500977
0.538719643217618 0.0282900035381317
1.53871964321762 0.0280499994754791
2.53871964321762 0.0278599917888641
3.53871964321762 0.0276499927043915
4.53871964321762 0.0273900032043457
5.53871964321762 0.0273499965667725
6.53871964321762 0.0273100018501282
7.53871964321762 0.0271099984645844
8.53871964321762 0.0272200047969818
21 0.0272200047969818
};
\addplot [ptamp, dashed, mark = triangle*, mark size = 2, mark options={solid}]
table {%
-44.4715803134222 0.892490000277758
-43.4715803134222 0.892589998990297
-42.4715803134222 0.891179999709129
-41.4715803134222 0.889560000598431
-40.4715803134222 0.887709999829531
-39.4715803134222 0.886490000784397
-38.4715803134222 0.884899999946356
-37.4715803134222 0.883940000087023
-36.4715803134222 0.880810000002384
-35.4715803134222 0.879930000752211
-34.4715803134222 0.874000000208616
-33.4715803134222 0.871569997817278
-32.4715803134222 0.867870000004768
-31.4715803134222 0.863209998607635
-30.4715803134222 0.857289999723434
-29.4715803134222 0.851909999549389
-28.4715803134222 0.844969998300076
-27.4715803134222 0.837019999325275
-26.4715803134222 0.825240002572536
-25.4715803134222 0.815850001573563
-24.4715803134222 0.799459998309612
-23.4715803134222 0.783410000801086
-22.4715803134222 0.764039999246597
-21.4715803134222 0.738259997963905
-20.4715803134222 0.711550003290176
-19.4715803134222 0.67654000222683
-18.4715803134222 0.638910004496574
-17.4715803134222 0.594249999523163
-16.4715803134222 0.53996000289917
-15.4715803134222 0.490780001878738
-14.4715803134222 0.435879999399185
-13.4715803134222 0.377209997177124
-12.4715803134222 0.325249993801117
-11.4715803134222 0.277140009403229
-10.4715803134222 0.231110000610352
-9.47158031342219 0.19436000585556
-8.47158031342219 0.164840000867844
-7.47158031342219 0.14028000831604
-6.47158031342219 0.122110003232956
-5.47158031342219 0.109889996051788
-4.47158031342219 0.0992500007152557
-3.47158031342219 0.0927100062370301
-2.47158031342219 0.0875900030136109
-1.47158031342219 0.0848800003528595
-0.471580313422193 0.0814899981021882
0.528419686577807 0.0791700005531311
1.52841968657781 0.0776300013065339
2.52841968657781 0.076559990644455
3.52841968657781 0.0749899923801423
4.52841968657781 0.0748899936676025
5.52841968657781 0.0739700019359588
21  0.0739700019359588
};
\end{axis}
\end{tikzpicture}}
	\caption{Classification error rate of \sinfoni\ on the MNIST validation dataset for different rate/channel uses constraints as a function of normalized SNR.} %
	\label{fig:sem_mnist_ntx}
\end{figure}
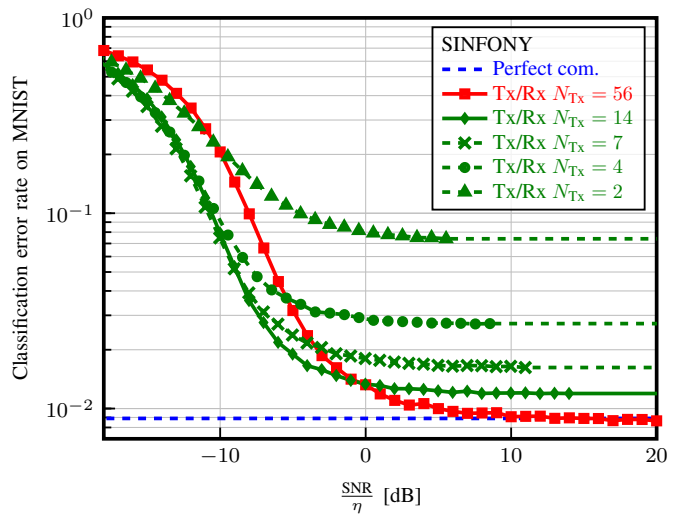

Since one of the main advantages of semantic communication lies in savings of information rate, we also investigate the influence of the number of channel uses $\ntx$ on MNIST classification error rate shown in Fig.~\ref{fig:sem_mnist_ntx}. From a practical point of view, we fix the information bottleneck by the output dimension $\ntx$ and maximize the mutual information $\mi[\params]{\relvec;\yvec}$. Decreasing the number of channel uses from $\ntx=14$ to $2$ and accordingly the upper bound $\mic$ on the mutual information $\mi[\params]{\semvec;\yvec}$, i.e., compression rate, from \eqref{eq:mi_channel_capacity} or \eqref{eq:entropy_independent}, we observe that the error floor at high SNR increases. We assume that, since the channel capacity decreases with SNR and $\ntx$, higher compression is required for reliable transmission through the channel in the training SNR interval. For $\ntx=56$, almost no error floor occurs at the cost of a smaller channel encoding gain.  This means compression and channel coding are balanced based on the channel condition, i.e., training SNR region, to find the optimal trade-off to maximize $\mi[\params]{\relvec;\yvec}$, which we can also observe in unshown simulations.

\subsubsection{Semantic vs. Classic Design}
\label{sec:semvsclassic}

\begin{figure}[!t]
	\centerline{\begin{tikzpicture}%
\begin{axis}[
width=10.5cm,
scale=0.7,
scale only axis,
xlabel={$\frac{\snr}{\speceff}$ [dB]}, %
xmajorgrids,
xmin=-17, xmax=21,
xminorgrids,
ylabel={Classification error rate on MNIST},
ymajorgrids,
ymin=3.5e-3, ymax=1,
yminorgrids,
ymode=log,
axis background/.style={fill=white},
legend style={at={(0.35, 0.69)}, anchor=west},
legend entries={
				{\color{\clsd} \centralimagecomm},
				{\hspace{-1.075cm} \sinfoni},
				{\color{\cloamp}\perfcomm},
				{\color{\clcmd}\txrx\ $\ntx=56$},
				{\color{\clsd}Classic digital com.},
				{\color{\clsd}\hspace{0.2cm}-$\rc$$=$$0.25$, BPSK},
				{\color{\clsd}\hspace{0.2cm}-$\rc$$=$$0.50$, BPSK},
				{\color{\clsd}\hspace{0.2cm}-$\rc$$=$$0.25$, 16QAM},
				{\color{\cldetnet}Analog semantic AE},
				},
]
\addplot [ptsd, mark=x, mark size=3, mark options={solid}]
table {%
-20 0.90
17.5258751031689 0.90204
18.0258751031689 0.8696
18.5258751031689 0.00885999999999998
19.0258751031689 0.00400000810623169
25 0.00400000810623169
};
\addplot [ptoamp, empty legend, dashed, mark = None, mark options={solid}]
table {%
0 0
1 2
};
\addplot [ptoamp, dashed, mark = None, mark options={solid}]
table {%
-20 0.008899986743927
25 0.008899986743927
};
\addplot [ptcmd]
table {%
-30 0.862450003623962
-29 0.858670000731945
-28 0.853880000114441
-27 0.84428000152111
-26 0.834960001707077
-25 0.826350000500679
-24 0.815559999644756
-23 0.799390000104904
-22 0.784559999406338
-21 0.763169999420643
-20 0.739959993958473
-19 0.712070000171661
-18 0.680809995532036
-17 0.639479997754097
-16 0.594949996471405
-15 0.541249999403954
-14 0.479689997434616
-13 0.410810005664825
-12 0.345440012216568
-11 0.270700007677078
-10 0.205979990959168
-9 0.144449996948242
-8 0.099260002374649
-7 0.0663699924945831
-6 0.0446700036525727
-5 0.0317500054836273
-4 0.0236199855804443
-3 0.0186600029468537
-2 0.0161800026893616
-1 0.014139997959137
0 0.0132299959659576
1 0.0118799984455109
2 0.0110099911689758
3 0.0104199945926666
4 0.0105999946594239
5 0.0099900007247925
6 0.0096700012683868
7 0.00945001244544985
8 0.00949000120162968
9 0.00952000021934507
10 0.00908000469207759
11 0.00912000536918645
12 0.00914000272750859
13 0.00891000628471372
14 0.00896001458168028
15 0.00890001058578493
16 0.00888999700546267
17 0.00867000818252561
18 0.00879999995231628
19 0.00879001021385195
20 0.00865001082420347
25 0.00865001082420347
};
\addplot [ptoamp, empty legend, dashed, mark = None, mark options={solid}]
table {%
0 0
1 2
};
\addplot [ptsd]
table {%
-20 0.90
13.2514762367513 0.89876
13.7514762367513 0.86821
14.2514762367513 0.0127200000000001
14.7514762367513 0.0091
25 0.0091
};
\addplot [ptsd, dashed, mark = diamond*, mark options={solid}, mark size = 2]
table {%
-20 0.90
14.2514762367513 0.89514
14.7514762367513 0.5805
15.2514762367513 0.0091
25 0.0091
};
\addplot [ptsd, dashed, mark = star, mark options={solid}, mark size = 3]
table {%
-20 0.90
14.7514762367513 0.9014
15.2514762367513 0.87649
15.7514762367513 0.0370900000000001
16.2514762367513 0.00909999999999989
25 0.0091
};
\addplot [ptdetnet, mark = triangle]
table {%
-30 0.85943
-29 0.85576
-28 0.84991
-27 0.84082
-26 0.83209
-25 0.81962
-24 0.80949
-23 0.79453
-22 0.77796
-21 0.75997
-20 0.73895
-19 0.71074
-18 0.68015
-17 0.64297
-16 0.60093
-15 0.55676
-14 0.5041
-13 0.44828
-12 0.38873
-11 0.32663
-10 0.2638
-9 0.20959
-8 0.15686
-7 0.11428
-6 0.0837
-5 0.0590400000000001
-4 0.0434599999999999
-3 0.03146
-2 0.02504
-1 0.0205
0 0.01732
1 0.01524
2 0.0140100000000001
3 0.01274
4 0.01191
5 0.01123
6 0.0108699999999999
7 0.0104299999999998
8 0.0103899999999999
9 0.0100299999999999
10 0.01007
11 0.0099499999999999
12 0.0100999999999999
13 0.0097799999999999
14 0.00975999999999999
15 0.00976999999999995
16 0.00975000000000004
17 0.00988000000000011
18 0.00980999999999987
19 0.00983000000000001
20 0.00980000000000003
25 0.00980000000000003
};
\end{axis}
\end{tikzpicture}}
	\caption{Classification error rate of \sinfoni\ with different kinds of optimized Tx/Rx modules and central image processing with digital image transmission on the MNIST validation dataset as a function of normalized SNR.} %
	\label{fig:sem_mnist_classic}
\end{figure}
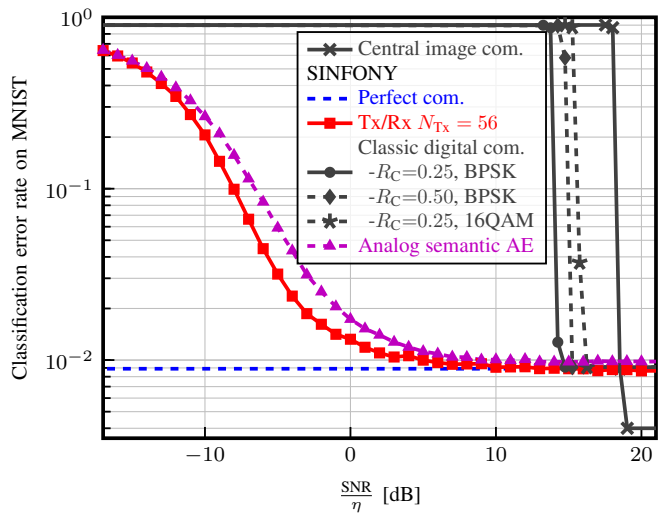

Finally, we compare semantic and classic communication system designs. For the classic digital design, we first assume that the images are compressed lossless and protected by a channel code for transmission and reliable overall image classification by the central unit based on $\aprob_{\rxpars}(\relvec|\semvec)$  (\centralimagecomm). We apply Huffman encoding to a block containing $100$ images $\semvecreal_{\indi}$ where each RGB color entry contains $8$ bits.

For fairness, we also compare to a \sinfoni\ version where Tx and Rx modules of Tab.~\ref{tab2:com_sem} are replaced by a classic design (\sinfoniclassdig): We first compress each element of the feature vector $\featvec_{\indi}$ that is computed in $32$-bit floating-point precision in the distributed setting \sinfonichan\ to $16$-bit. Then, we apply Huffman encoding to a block containing $100$ feature vectors of length $\Nfeat$.

Further, we use a 5G LDPC channel code implementation from \cite{hoydis_sionna_2022} with interleaver, rate $\rc=\{0.25,0.5,0.25\}$ and long block length of $\{15360,16000,15360\}$, and modulate the code bits with \{BPSK, BPSK, 16-QAM\}. For digital image transmission, we use rate $\rc=0.25$ with block length of $15360$ and BPSK modulation. At the receiver, we assume belief propagation decoding, where the noise variance is perfectly known for LLR computation.

The results in Fig.~\ref{fig:sem_mnist_classic} reveal tremendous information rate savings for the semantic design with \sinfoni: We observe an enormous SNR shift of roughly $20$ dB compared to the classic digital design w.r.t. to both image (\centralimagecomm) and feature transmission (\sinfoniclassdig). Note that the classic design is already near the Shannon limit and even if we improve it by ML we are only able to shift its curve by a few dB. The reason may lie in overall system optimization with \sinfoni\ w.r.t. semantics and analog encoding of $\xvec$.

\subsubsection{\sinfoni\ vs. Analog ``Semantic" Autoencoder}
\label{sec:semvsae}

To distinguish both influences, we also implemented the approach of \eqref{eq:jscc} according to Shannon by analog AEs. The analog AE has been introduced by O'Shea and Hoydis in \cite{oshea_introduction_2017}. From the viewpoint of semantic communication, it resembles the semantic approach from \cite{farsad_deep_2018, bourtsoulatze_deep_2019, xie_deep_2021, beck_swarm_2023} without differentiating between semantic and channel coding and the mutual information constraint $\mi{\xvec;\yvec}$ like in \cite{xie_deep_2021}. We trained the AE matching the Tx and Rx module in Tab.~\ref{tab2:com_sem} with mean square error criterion for reliable transmission of the feature vector $\featvec$
with \sinfoni\ training settings. %
The Rx module consists of one ReLU layer of width $\nrx=\ntx$ providing the estimate of $\featvec$.
We provide results (\sinfoniclassanae) in Fig.~\ref{fig:sem_mnist_classic}:
Indeed, most of the shift is due to analog encoding. By this means, we further avoid the typical thresholding behavior of a classic digital system seen at $14$ dB.

In conclusion, this surprisingly clear result justifies an analog ``semantic" communications design and shows its huge potential to provide bandwidth savings. However, introducing the semantic \rv\ $\relvec$ by \sinfoni, we can further shift the curve by $2$ dB and avoid a slightly higher error floor compared to the analog ``semantic" AE. With expect a larger performance gap with more challenging image datasets such as CIFAR10. More importantly, the main benefit of \sinfoni\ lies in lower training complexity: We avoid separate and possibly iterative semantic and communication training procedures. %

\section{Conclusion}
\label{sec:conclusion}

Motivated by the approach of Bao, Basu et al. \cite{bao_towards_2011, basu_preserving_2014} and inspired by Weaver's notion of semantic communication \cite{weaver_recent_1949}, we brought the terminus of a semantic source to the context of communications by considering its complete Markov chain. We defined the task of semantic communication in the sense of a data-reduced and reliable transmission of communications sources / messages over a communication channel such that the semantic Random Variable (\rv) at a recipient is best preserved. We formulated its design either as an Information Maximization or as an Information Bottleneck optimization problem covering important implementations aspects like the reparametrization trick and solved the problems approximately by minimizing the cross entropy that upper bounds the negative mutual information. With this article, we distinguish from related literature \cite{bao_towards_2011, basu_preserving_2014, xie_deep_2021, shao_learning_2022, beck_swarm_2023} in both classification and perspective of semantic communication and a different ML-based solution approach.

Finally, we proposed the ML-based semantic communication system \sinfoni\ for a distributed multipoint scenario: \sinfoni\ communicates the meaning behind multiple messages that are observed at different senders to a single receiver for semantic recovery. We analyzed \sinfoni\ by processing images as an example of messages. Notably, numerical results reveal a tremendous rate-normalized SNR shift up to $20$ dB compared to classically designed communication systems.

\bibliographystyle{IEEEtran}
\bibliography{IEEEabrv, references}

\begin{IEEEbiography}[{\includegraphics[width=1in,height=1.25in,clip,keepaspectratio]{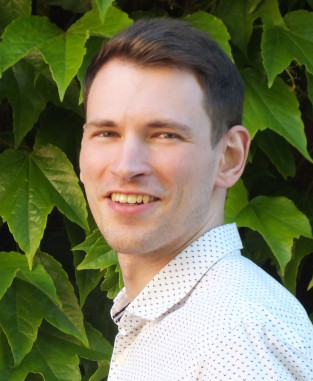}}]{Edgar Beck}
	(Graduate Student Member, IEEE) received the B.Sc. and M.Sc. degrees in electrical engineering from the University of Bremen, Germany, in 2014 and 2017, respectively, where he is currently pursuing a Ph.D. degree in electrical engineering at the Department of Communications Engineering (ANT). His research interests include cognitive radio, compressive sensing, massive MIMO systems, semantic communication, and machine learning for wireless communications.
	
	Edgar Beck %
	was a recipient of the OHB Award for the best M.Sc. degree in Electrical Engineering and Information Technology at the University of Bremen in 2017.
\end{IEEEbiography}

\begin{IEEEbiography}[{\includegraphics[width=1in,height=1.25in,clip,keepaspectratio]{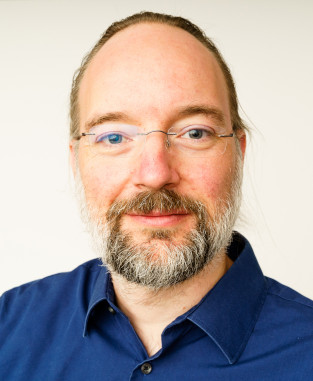}}]{Dr. Carsten Bockelmann}
	(Member, IEEE) received Dipl.-Ing. and Ph.D. degrees in electrical engineering from the University of Bremen, Germany, in 2006 and 2012, respectively. Since 2012, he has been a Senior Research Group Leader with the University of Bremen, coordinating research activities regarding the application of compressive sensing and machine learning to communication problems. His research interests include massive machine-type communication, ultra-reliable low latency communications and industry 4.0, compressive sampling, and channel coding.
\end{IEEEbiography}

\begin{IEEEbiography}[{\includegraphics[width=1in,height=1.25in,clip,keepaspectratio]{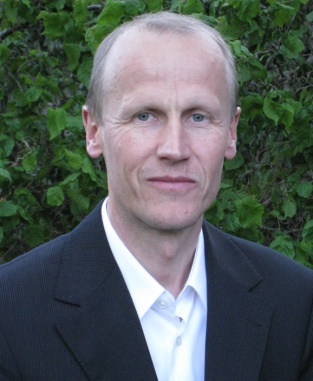}}]{Prof. Dr. Armin Dekorsy}
	(Senior Member, IEEE) received his B.Sc. from the University of Applied Sciences Konstanz, his M.Sc. from the University of Paderborn, and his Ph.D. from the University of Bremen in 2000, all in communications engineering. He is distinguished by eleven years of industry experience in leading research positions, notably as DMTS at Bell Labs and as Research Coordinator Europe Qualcomm, and by conducting more than 50 (inter)national research projects. He is head of the Department of Communications Engineering, founder and managing director of the Gauss-Olbers Space-Technology-Transfer-Center, and has been a board member of the Technology Center for Informatics and Information Technology (TZI) for the last 10 years. His current research focuses on distributed signal processing, compressive sensing, information bottleneck method, semantic communications, and machine learning, leading to the development of technologies for 5G/6G, industrial radio, and LEO/GEO satellite communications. He is a Senior Member of the IEEE Communications and Signal Processing Society and a member of the VDE/ITG Expert Committee “Information and System Theory”. He is also co-author of the textbook ``Nachrichtenübertragung, Release 6, Springer`, a bestseller in the German-speaking world on communication technologies.
\end{IEEEbiography}

\end{document}